# Dirac e il monopolo magnetico

Anna Maria Aloisi

IPSIA "A.Meucci", Cagliari, http:apmf.interfree.it, ampf@interfree

Pier Franco Nali

*Servizio per l'Innovazione Tecnologica e per le Tecnologie dell'Informazione e delle Comunicazioni,*

*Regione Sardegna*

LA FISICA, più di ogni altro settore della scienza contemporanea, incarna il mito della grande impresa scientifica. Nei grandi laboratori mondiali, come il CERN e il Fermilab, si concentrano infatti ingenti risorse e investimenti giganteschi, con il concorso di più nazioni. Ed è fuor di dubbio che questo modello di organizzazione industriale della ricerca, per il quale è stato coniato il termine *big science*, può vantare successi straordinari. Ma nello stesso tempo la ricerca d'avanguardia, condotta in questi laboratori ipertecnologici, inevitabilmente patisce i limiti di quella stessa tecnologia che ne costituisce il punto di forza, e lascia vaste aree non indagate. L'esplorazione di queste regioni, inaccessibili alla sperimentazione diretta, è un'affascinante avventura intellettuale: un'arena smisurata in cui si confrontano teorie speculative libere da vincoli che non siano quelli imposti dal rigore matematico. In effetti, le teorie fondamentali che stanno alla base delle nostre conoscenze sul mondo fisico sono nate come teorie speculative, che solo successivamente, dopo aver trovato le necessarie conferme sperimentali, sono state accettate universalmente. Semplificando, possiamo definire le teorie speculative come catene di implicazioni, costruite con i mezzi della logica e della matematica pura assumendo come presupposti un numero limitato di principi fisici fondamentali desunti per generalizzazione da fatti sperimentali d'immediata evidenza. Naturalmente una teoria speculativa che si candida a diventare teoria fondamentale deve spiegare i fatti già noti, ma, quando il formalismo matematico viene sviluppato in tutta la sua generalità, senza imporre restrizioni arbitrarie, la teoria deve essere anche in grado di produrre previsioni verificabili di nuovi fatti, che ne decreteranno il destino definitivo.

L'esempio più emblematico è quello della teoria della relatività generale di Einstein, costruita intorno al 1915 sulla base del principio di equivalenza con un gigantesco sforzo matematico. Molte previsioni di questa teoria, che rappresenta certamente una delle massime conquiste intellettuali dell'umanità, sono state confermate con esperimenti accurati, mentre altre sono in corso di verifica con esperimenti ancora più precisi. Un approccio molto simile, basato su considerazioni di consistenza e di eleganza matematica, fu introdotto da Dirac verso la fine degli anni '20 nella teoria quantistica, producendo le sorprendenti previsioni dell'energia negativa e delle antiparticelle, meravigliosamente confermate alcuni anni dopo.

Nel 1931 Dirac avanzò una nuova affascinante previsione sulla linea di quelle prime ricerche pionieristiche: quella della possibile esistenza di monopoli magnetici. Si tratta di un'idea mai confermata sperimentalmente, e per tale motivo a lungo accantonata, ma che negli ultimi decenni, con l'apertura all'esplorazione di nuovi domini energetici, è tornata in auge, ispirando ricerche d'avanguardia in fisica e in cosmologia. L'idea dei monopoli magnetici ha anche un particolare interesse storico e didattico, per il modo magistrale in cui venne presentata originariamente da Dirac e per gli sviluppi che produsse, ed è proprio questo interesse che ci ha spinti alla stesura di questo articolo.

## *Come due gemelli*

Il problema della dualità incompleta tra elettricità e magnetismo ha sempre affascinato i fisici. È noto fin dall'antichità che i poli di un magnete non possono venire separati. I primi sperimentatori osservarono che se si divide un magnete, i suoi frammenti sono altrettanti magneti completi, ciascuno con due poli. Con l'introduzione del concetto di campo verso la metà del XIX secolo, quest'osservazione apparentemente elementare fu descritta rappresentando il campo d'azione di un magnete come un insieme di linee di forza chiuse che riempiono lo spazio. I campi magnetici prodotti artificialmente, come quello di una spira conduttrice percorsa da corrente, sono in tutto simili ai campi dei magneti naturali, ed in particolare hanno linee di forza chiuse. Questa proprietà delle linee di forza magnetiche si manifesta in tutte le situazioni note e costituisce un carattere distintivo fondamentale del magnetismo, descritto matematicamente scrivendo $rot\mathbf{H} \neq 0$. Qualsiasi magnete, naturale o artificiale, si comporta perciò come un dipolo, un sistema composto da due poli di polarità opposte indicate convenzionalmente come nord e sud: le linee di forza che fuoriescono dal polo nord s'incurvano e si richiudono su se stesse attraverso il polo sud. La descrizione matematica di tale comportamento, caratterizzato dal fatto fondamentale che le linee di forza non divergono, è fornita dall'equazione $div\mathbf{H} = 0$, che esprime la condizione di assenza di sorgenti per il campo magnetico. Un campo senza sorgenti viene detto *solenoidale*.

A differenza del campo magnetico il campo elettrico non è solenoidale. Inoltre, in condizioni stazionarie, è *irrotazionale*, cioè non ha linee di forza chiuse. Questa proprietà del campo elettrico viene espressa dall'equazione $rot\mathbf{E} = 0$. In un dipolo elettrico, i cui poli sono due cariche di segno opposto, le linee di forza fuoriescono dal polo positivo e convergono nel polo negativo. Le cariche che costituiscono un dipolo elettrico possono venire separate, e l'esistenza di cariche isolate è un carattere distintivo del campo elettrico. Le linee di forza che fuoriescono da una carica isolata si irradiano nello spazio con una distribuzione isotropa. Ciò si esprime matematicamente scrivendo $div\mathbf{E} \neq 0$ ovvero, ricorrendo al teorema di Gauss, con l'equazione $\Phi_E = 4\pi q$. La carica $q$ che compare in quest'equazione è la sorgente del campo elettrico; la si potrebbe definire un "monopolo elettrico". Pur con queste differenze, se vengono osservati da grande distanza e lontano dalle sorgenti, i campi di dipolo, elettrico e magnetico, appaiono identici.

È naturale porsi la seguente domanda: questa incompleta simmetria tra elettricità e magnetismo è un aspetto fondamentale della Natura, o soltanto un'apparenza dietro cui si nascondono situazioni non ancora svelate in cui $div\mathbf{H} \neq 0$?

Supponiamo che esista un polo magnetico isolato in un punto fisso dello spazio. Il suo campo magnetostatico irradia nello spazio linee di forza esattamente uguali a quelle del campo elettrostatico di una carica isolata. Un tale ipotetico polo isolato, o monopolo, verrà perciò osservato come una particella dotata di carica magnetica, cioè con polarità magnetica nord o sud. Il flusso magnetico uscente da una superficie chiusa che racchiude una di queste cariche magnetiche sarebbe dato da $\Phi_H = 4\pi g$, dove $g$ rappresenta la carica magnetica del monopolo[1]. Tuttavia la corrispondenza tra il campo magnetostatico di un monopolo e il campo elettrostatico di una particella carica non è perfetta. Il campo elettrostatico $\mathbf{E} = \frac{q}{r^2}\hat{\mathbf{r}}$, essendo irrotazionale, deriva infatti da un potenziale scalare con un valore finito in ogni punto dello spazio, tranne nell'origine[2]. Il campo magnetostatico di un monopolo, pur essendo descritto dall'analoga equazione $\mathbf{H} = \frac{g}{r^2}\hat{\mathbf{r}}$, deriva invece da un potenziale vettore magnetico **A**, infinito (singolare) in tutti i punti di una semiretta che parte dal monopolo[3]. Tale semiretta singolare e il monopolo alla sua estremità, prendono il nome, rispettivamente, di *stringa di Dirac* e *monopolo di Dirac*. In generale una singolarità è di difficile trattazione matematica e di ambigua interpretazione fisica: è un campanello d'allarme che qualcosa nella teoria non va. Si potrebbe sostenere che la singolarità del potenziale vettore è fisicamente ininfluente, essendo il potenziale nient'altro che un artefatto matematico utilizzato nel calcolo dei campi, che sono le sole entità con un significato fisico diretto. Quest'osservazione è senz'altro valida nel mondo macroscopico descritto dalla fisica classica, ma non è accettabile in un contesto quantistico. Infatti la funzione d'onda di una particella carica in un campo elettromagnetico, che contiene tutta l'informazione sullo stato del sistema, è la soluzione di un'equazione d'onda in cui entrano i potenziali elettromagnetici (per le particelle con spin vi entrano anche i campi). Pertanto *un sistema quantistico viene influenzato dal potenziale vettore magnetico*[4], la cui singolarità porterebbe a risultati fisici privi di senso. Si deve allora assumere la congettura, molto forte, che la singolarità del potenziale vettore nella stringa di Dirac che accompagna il monopolo non sia fisicamente osservabile, ossia non eserciti nessuna influenza sul mondo esterno. Tale singolarità deve risiedere in un certo senso in una dimensione extra (un subspazio). Per questa ragione, e per certe altre loro proprietà piuttosto inconsuete, alcuni fisici non credono nell'esistenza dei monopoli e li considerano un mero prodotto della fantascienza[5].

Si può dimostrare facilmente la singolarità del potenziale vettore magnetico nella stringa di Dirac calcolandone la circuitazione intorno alla circonferenza *C* (v. *Fig. 1*). Il campo ha una simmetria radiale intorno al monopolo *M*. Si può scegliere a piacere un punto *P* sulla superficie che

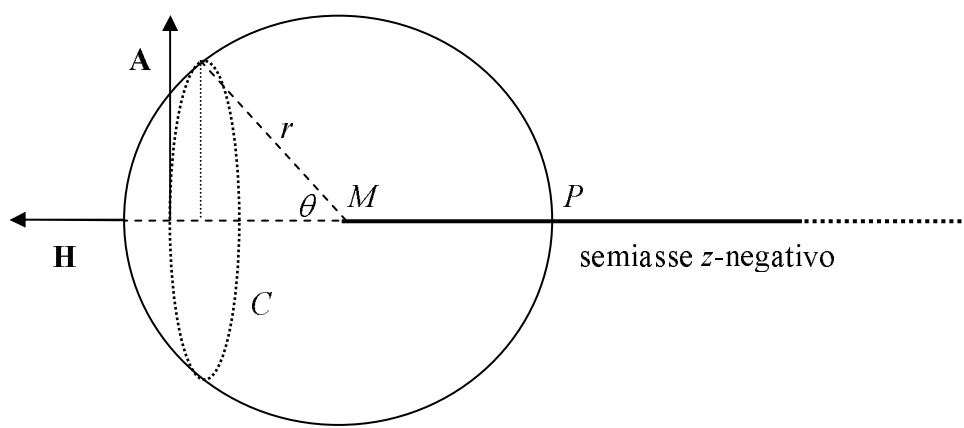

*Fig. 1*

delimita una regione sferica di raggio *r* che circonda il monopolo, e tracciare da *M* la semiretta *MP*

che facciamo convenzionalmente coincidere col semiasse $z$-negativo. $C$ viene percorsa in senso orario dal vettore **A**, che ha una simmetria intorno a quest'asse. La circuitazione di **A** intorno a $C$ è $\oint_C \mathbf{A} \cdot \mathbf{ds} = A_\varphi \cdot 2\pi r \sin\theta$, ed è uguale al flusso $\Phi_H = \oiint H dS = H_r S = \frac{g}{r^2} S$ uscente dalla regione sferica delimitata a destra da $C$, di raggio $r$ ed area $S = \int_0^{2\pi}\int_0^\theta r^2 \sin\theta d\theta d\phi = 2\pi r^2(1-\cos\theta)$.

Otteniamo pertanto $2\pi r A_\phi \sin\theta = \frac{g}{r^2} \cdot 2\pi r^2(1-\cos\theta) = 2\pi g(1-\cos\theta)$, ed infine

$$A_\phi = \frac{g}{r}\frac{(1-\cos\theta)}{\sin\theta},$$

che è infinito (singolare) per $\theta = \pi$, cioè sul semiasse $z$-negativo.

Il potenziale **A** è dunque singolare su una semiretta (nel nostro esempio il semiasse $z$-negativo $MP$ della *Fig. 1*) con una posizione che dipende dalla scelta (arbitraria) di una direzione e che ha l'origine nel monopolo $M$. Tale semiretta singolare è appunto la stringa di Dirac.

Una raffigurazione efficace del monopolo e della stringa di Dirac che l'accompagna è fornita da un lungo solenoide rettilineo di piccolo diametro. Le linee di forza si irradiano all'esterno dalle estremità del solenoide, assumendo una distribuzione isotropa a grandi distanze. In tal modo le due estremità appaiono a grandi distanze come una coppia di monopoli magnetici di polarità opposta (polo$N$=monopolo, polo$S$=antimonopolo) (v. *Fig. 2*).

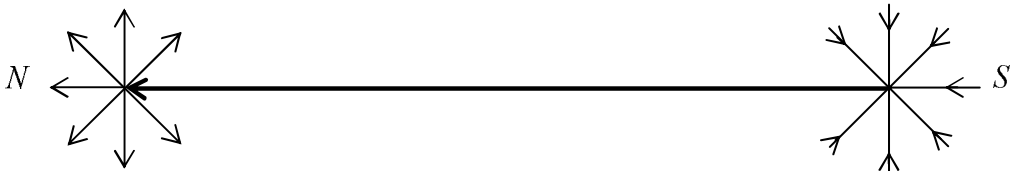

*Fig. 2*

Nel caso limite di un semi-solenoide infinito di diametro nullo, ottenuto allontanando un'estremità (p. es. $S$) all'infinito, l'estremità che rimane accessibile ($N$) apparirebbe come un monopolo isolato, e il semi-solenoide infinito come la stringa di Dirac ad esso associata. Affinché il flusso magnetico non svanisca quando il diametro si annulla, il campo all'interno del semi-solenoide deve diventare infinito. Un campo infinito all'interno richiede una corrente circolante nelle spire infinita o, in alternativa, una densità lineare di spire infinita (vale a dire, nessuna separazione tra due spire contigue qualsiasi infinitamente sottili).

Per una trattazione matematica, calcoliamo dapprima il flusso magnetico all'interno di un semi-solenoide rettilineo infinito di sezione circolare di semi-diametro $a$. Ragioniamo in prima approssimazione, considerando il campo irradiato a grandi distanze ($r \gg a$) come un campo radiale isotropo che segue una legge $1/r^2$ all'esterno del semi-solenoide, e un campo longitudinale uniforme all'interno. Successivamente determineremo il flusso nella stringa di Dirac con un passaggio al limite per $a \to 0$. Quando si giunge al limite di diametro nullo, l'approssimazione della distribuzione isotropa delle linee di forza a grandi distanze diviene valida a tutte le distanze.

Il flusso $\Phi_H = \iint H dS$ attraverso una sezione $S$ del semi-solenoide di semi-diametro $a$ è uguale alla circuitazione $A_\phi \cdot 2\pi a = 2\pi g(1-\cos\theta) = 2\pi g(1 - \frac{z}{\sqrt{z^2+a^2}})$ del potenziale vettore intorno alla circonferenza del semi-solenoide. A grandi distanze ($|z| \gg a$) possiamo assumere che il campo longitudinale all'interno del semi-solenoide sia uniforme e scrivere:

$$\Phi_H = \iint H dS \simeq H_z \cdot S = H_z \cdot \pi a^2 = A_\phi \cdot 2\pi a = 2\pi g(1 - \frac{z}{\sqrt{z^2+a^2}}) ,$$

da cui ricaviamo per il campo all'interno:

$$H_z \simeq 2\pi g(1 - \frac{z}{\sqrt{z^2+a^2}}) \cdot \frac{1}{\pi a^2} = 8g(1 - \frac{z}{\sqrt{z^2+a^2}}) \cdot \left(\frac{1}{2a}\right)^2 .$$

La successione discontinua $d_a(\rho) = \begin{cases} \frac{1}{2a}, \rho \in (-a,+a) \\ 0, \rho \notin (-a,+a) \end{cases}$, dove $\rho = \pm\sqrt{x^2+y^2}$, che compare nella formula appena ricavata per $H_z$, tende alla distribuzione delta di Dirac $\delta(\rho)$ per $a \to 0$ [6].

Otteniamo pertanto per il campo (longitudinale) nella stringa di Dirac ($x=y=0$, $z<0$):

$$H_z(\text{int}) = \lim_{a \to 0}\left[8g(1 - \frac{z}{\sqrt{z^2+a^2}}) \cdot \left(\frac{1}{2a}\right)^2\right] = 8g(1 - \frac{z}{|z|}) \cdot \lim_{a \to 0}\left[d_a(\rho)\right]^2\bigg|_{\rho=0} = 16g\left[\delta(\rho)\right]^2\bigg|_{\rho=0} = 16g\left[\delta(\sqrt{x^2+y^2})\right]^2\bigg|_{x=y=0} = \infty \text{ [7]},$$

e per il flusso corrispondente: $\Phi_H(\text{int}) = \lim_{a \to 0} \iint H dS = \lim_{a \to 0} 2\pi g(1 - \frac{z}{\sqrt{z^2+a^2}}) = 2\pi g(1 - \frac{z}{|z|}) = 4\pi g$.

Il campo $H_M$ all'estremità di un semi-solenoide infinito (nell'origine $M$ del semiasse $z$-negativo nella *Fig. 1*) è legato alla corrente che circola nelle spire dalla relazione $H_M = \frac{4\pi}{c} \cdot \frac{1}{2} nI = \frac{4\pi}{c} \cdot \frac{1}{2} I/l$ (dove $n$ è il numero di spire per unità di lunghezza, $l$ la separazione tra le spire supposte equidistanziate e $I$ la corrente), mentre $H_z$ (sull'asse del semi-solenoide) è legato a $H_M$ dalla relazione

$$H_z = H_M(1 - \frac{z}{\sqrt{z^2+a^2}}) \text{ [8]}.$$

A grandi distanze ($|z| \gg a$) $H_z = 2H_M = \frac{4\pi}{c} \cdot nI$ per $z<0$ e $H_z = 0$ per $z>0$. Passando al limite per $a \to 0$ otteniamo per la corrente:

$$I = \frac{4c}{\pi} gl \left[\delta(\sqrt{x^2+y^2})\right]^2\bigg|_{x=y=0} = \infty$$

e, in alternativa, per la densità lineare delle spire:

$$n = \frac{4c}{\pi} \frac{g}{I} \left[\delta(\sqrt{x^2+y^2})\right]^2\bigg|_{x=y=0} = \infty .$$

Nel passaggio al limite per $a \to 0$ ($\theta \to \pi$) la corrente diventa dunque infinita.

Il flusso $\Phi_H(\text{int})$ trasportato attraverso la stringa di Dirac "nel monopolo" si mantiene invece finito, uguale a $4\pi g$, e fornisce l'intero flusso irradiato dal monopolo. Da dove proviene questo flusso? Evidentemente dall'antimonopolo che si trova all'infinito! Quindi possiamo dire che un monopolo, per quanto isolato, è sempre accoppiato con un antimonopolo.

Il flusso magnetico all'interno di un solenoide reale resta "nascosto" al mondo esterno fino a quando non riappare e viene irradiato alle estremità; ma sappiamo che all'interno il flusso esiste realmente, sicché il bilancio complessivo interno-esterno rimane in parità, e continua a valere la legge $div\mathbf{H} = 0$ (condizione di assenza di sorgenti). A differenza di un solenoide reale, la stringa di Dirac, che non è di natura materiale, è inaccessibile e trasporta un flusso del tutto inosservabile finché non viene irradiato all'esterno, dove riappare come un flusso netto $\Phi_H = 4\pi g$ in violazione della legge della divergenza ($div\mathbf{H} \neq 0$). Il monopolo si manifesta quindi come una sorgente di carica magnetica *g*, che irradia nello spazio il flusso magnetico trasportato attraverso la stringa di Dirac in un canale subspaziale.

La stringa di Dirac ha dunque la funzione di chiudere le linee di forza tra due poli magnetici isolati attraverso una specie di canale invisibile. Questo "passaggio segreto" è il meccanismo che fa apparire completa la simmetria tra elettricità e magnetismo senza richiedere modifiche delle leggi fondamentali. Ma sfortunatamente introduce nella teoria elettromagnetica una singolarità[9], ancor più strana perché inosservabile e inaccessibile, cioè non fisica.

In virtù di questa simmetria l'elettricità e il magnetismo ci appaiono come due gemelli, ma solo se li osserviamo da rispettosa distanza!

### *Il programma di Dirac*

Verso la fine del XIX secolo Poincaré aveva analizzato, nel quadro della teoria elettromagnetica classica, il problema del moto di una particella carica nel campo magnetico di un polo isolato. In questo tipo di ricerche l'esistenza del monopolo veniva assunta *a priori* per via ipotetica. Dirac si pose un obiettivo molto più ambizioso: far emergere l'esistenza della nuova particella *a posteriori*, come conseguenza matematica della teoria quantistica.

Nell'introduzione del suo articolo del 1931[10] sul monopolo, Dirac si dilungava sul rapporto tra fisica e matematica, annunciando un vero e proprio programma di ricerca per lo sviluppo della teoria quantistica su basi matematiche. In un precedente articolo del 1928 egli aveva avanzato la previsione dell'esistenza di valori di energia negativa per l'elettrone e di una nuova particella, simile all'elettrone ma di carica elettrica opposta (positrone), che sarebbe stata scoperta da Anderson qualche anno più tardi (1932). La coerenza matematica della teoria era così perfetta che Dirac si convinse che le idee di energia negativa e antiparticella potevano essere accettate come plausibili, sebbene a quell'epoca non avessero ancora avuto conferme sperimentali. Ne concluse che le considerazioni basate sulla coerenza e l'eleganza matematica potevano essere usate per predire l'esistenza di una nuova particella. Era infatti convinto che fosse sempre più difficile sviluppare la fisica teorica in maniera diretta, ossia proponendo spiegazioni teoriche partendo dall'analisi dei dati sperimentali.

«Il costante progresso della fisica – scrive Dirac - richiede per la sua formulazione teorica una matematica sempre più avanzata. Questo è assolutamente naturale e prevedibile. Ciò che invece non era stato previsto dagli scienziati del secolo scorso era la particolare forma che avrebbe preso la linea di sviluppo della matematica; ci si aspettava cioè che la matematica sarebbe diventata sempre più complicata, ma sarebbe rimasta poggiata su una base permanente di assiomi e definizioni, mentre in realtà gli sviluppi della fisica moderna hanno richiesto una matematica che sposta continuamente i suoi fondamenti e diventa più astratta. [...] Sembra probabile che questo processo di crescente astrazione continuerà in futuro e che si dovrà associare il progresso della fisica ad una continua modificazione e generalizzazione degli assiomi alla base della matematica piuttosto che ad uno sviluppo logico di un qualunque schema matematico con fondamenti stabiliti definitivamente. Esistono oggi problemi fondamentali di fisica teorica che aspettano di essere risolti [...]. La soluzione di questi problemi richiederà probabilmente una revisione dei nostri concetti fondamentali ancor più drastica rispetto a qualsiasi altra fatta in precedenza. È molto probabile che questi cambiamenti saranno così grandi da porre al di là delle capacità dell'intelligenza umana la possibilità di ricavare le nuove idee necessarie direttamente dai tentativi di formulare in termini matematici i dati sperimentali. Il fisico teorico in futuro dovrà perciò procedere in maniera più indiretta. Il più potente metodo di progresso che possa oggi esser suggerito è quello dell'impiego di tutte le risorse della matematica pura nel tentativo di perfezionare e generalizzare il formalismo matematico che costituisce la base della fisica teorica, e dopo ogni successo ottenuto in questa direzione, tentare di interpretare le nuove acquisizioni matematiche in termini di entità fisiche».
Come si vede, il metodo suggerito privilegia la previsione teorica su basi matematico-deduttive, l'esperimento si rende necessario solo in una fase successiva di conferma.

Su tali presupposti Dirac avanzò una nuova originale idea «per molti aspetti paragonabile con quella delle energie negative»: l'esistenza di un quanto elementare di carica elettrica (corrispondente alla carica *e* di un elettrone) è strettamente connessa con l'esistenza di un quanto di carica magnetica. Dirac descrisse nell'ambito della meccanica quantistica le proprietà di un monopolo magnetico, e concluse che deve avere una carica magnetica quantizzata. Inoltre, quando un monopolo interagisce con una particella carica, deve esistere una relazione tra la carica elettrica della particella e la carica magnetica del monopolo. Ne consegue che anche la carica elettrica deve essere quantizzata. Questo fatto assume uno straordinario rilievo, perché se esistesse un monopolo magnetico con una carica magnetica opportuna, allora potrebbero esistere cariche elettriche frazionarie, con notevoli conseguenze in campo teorico: le cariche frazionarie dei quark, infatti, non hanno ancora trovato una spiegazione convincente.

La relazione che lega i quanti di elettricità e di magnetismo (chiamata condizione di quantizzazione di Dirac) è semplice ed elegante[11]:

$$(1)\, qg = \tfrac{1}{2}\hbar c \cdot n$$

Qual è il suo significato? Essa ci dice che l'esistenza dei monopoli magnetici non è in contrasto con i principi quantistici, a condizione che i quanti di carica elettrica e magnetica siano legati appunto dalla (1). La struttura matematica della teoria quantistica fornisce gli elementi per dedurre in modo naturale l'esistenza delle nuove particelle, senza che sia necessario introdurre nessun cambiamento nei principi quantistici fondamentali, ma semplicemente approfondendone le implicazioni. In un certo senso, i monopoli sono già implicitamente presenti nella struttura matematica della teoria quantistica ed è sufficiente semplicemente farli emergere. «L'oggetto del presente articolo – scrive Dirac con modestia – è mostrare che la meccanica quantistica in realtà non preclude l'esistenza di poli magnetici isolati». Ma poi prosegue: «Al contrario, l'attuale formalismo della meccanica quantistica, quando venga sviluppato in maniera naturale senza l'imposizione di restrizioni arbitrarie, porta inevitabilmente a delle equazioni d'onda la cui unica interpretazione fisica è il moto di un elettrone nel campo di un singolo polo. Questo nuovo sviluppo *non richiede alcun cambiamento* nel formalismo quando viene espresso nei termini di simboli astratti che denotano stati e osservabili, ma è una mera generalizzazione delle possibilità di rappresentazione di questi simboli astratti mediante funzioni d'onda e matrici. Sotto queste circostanze si dovrebbe essere sorpresi se la Natura non ne avesse fatto uso».

In breve: tutto ciò che è matematicamente consistente è anche fisicamente possibile. Se così non fosse Dirac ne stato sarebbe sorpreso. Ma come trasformò una tesi così ardita, e di carattere molto generale, nella sorprendente previsione dell'esistenza di monopoli magnetici?

*Ambigue fasi*

Per prima cosa, Dirac mise la fase della funzione d'onda sotto il suo formidabile microscopio intellettuale. Egli si pose più o meno la seguente domanda: se si segue una funzione d'onda attraverso un cammino chiuso, la fase torna al suo valore originale oppure no? Se non vi torna la fase è non-integrabile. E una fase non-integrabile è possibile in meccanica quantistica? Per seguire il suo ragionamento è necessario richiamare alcuni concetti basilari di meccanica quantistica.

Nella meccanica quantistica esiste un'ambiguità di principio nella definizione della funzione d'onda, che è sempre determinata a meno di un coefficiente costante (complesso) arbitrario. Se si esprime una funzione d'onda ordinaria, univocamente determinata, nella forma $\psi_0 = |\psi_0| \cdot e^{i\gamma}$, l'ambiguità si manifesta nella possibile aggiunta alla fase $\gamma$, variabile da punto a punto, di una costante arbitraria $\alpha$ che rende la funzione d'onda non univoca: $\psi = \psi_0 \cdot e^{i\alpha}$. In questo modo la fase della funzione d'onda in un punto $P$ qualsiasi può venire espressa in funzione del parametro $\alpha$: $\varphi_\alpha(P) = \gamma(P) + \alpha$. Poiché $\alpha$ non dipende da $P$ l'indeterminazione della fase è la stessa in tutti i punti per una stessa funzione d'onda. Di conseguenza, la fase in un particolare punto non ha un preciso significato fisico, solo la differenza di fase tra due punti qualsiasi è definita senza ambiguità. Questa assenza di univocità non può essere eliminata ma non costituisce una minaccia per la coerenza della teoria quantistica, perché non influisce su nessun risultato fisico e non introduce ambiguità in nessuna applicazione della teoria[12]. Potremo perciò mettere da parte il fattore di fase indeterminato $e^{i\alpha}$ e non considerarlo nel seguito della discussione.

Ma potremo invece andare oltre ed assumere, senza violare nessuna regola della meccanica quantistica, che neppure la differenza di fase tra due punti sia univocamente definita, a meno che i due punti siano contigui. Se i punti sono distanti la differenza di fase in generale dipende dalla curva che li congiunge. È il punto chiave di tutto il ragionamento! *Se si percorrono differenti curve che congiungono due punti dati si troveranno in generale differenze di fase diverse sulle diverse curve.* Se per esempio, con riferimento alla *Fig. 3*, andiamo da $A$ a $B$ lungo la curva $C_1$ otterremo una differenza di fase, diciamo $\Delta\varphi_1$, mentre percorrendo $C_2$ troveremo una differenza di fase $\Delta\varphi_2$, che sarà in generale diversa dalla precedente. Se percorriamo una curva chiusa il cambiamento totale della fase intorno a essa in generale non sarà nullo. Restando al nostro esempio, se andiamo da $A$ a $B$ lungo $C_1$ e quindi facciamo ritorno ad $A$ lungo $C_2$ troveremo una differenza di fase non nulla $\beta = \Delta\varphi_1 - \Delta\varphi_2$ per la curva chiusa $C = C_1 - C_2$.[13]

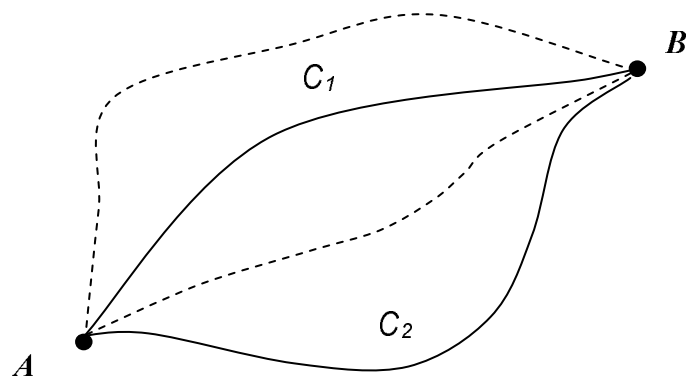

*Fig. 3*

Per affrontare matematicamente il problema in modo generale esprimiamo la fase della funzione d'onda in un qualsiasi punto $P$ come $\varphi_{\alpha,C}(P) = \gamma(P) + \beta_C + \alpha$, cioè come la somma del termine monodromo $\gamma$ (a un sol valore per ogni punto $P$), dell'indeterminazione $\beta$ (avente in generale un diverso valore per ogni curva chiusa $C$ passante per $P$) e della costante arbitraria $\alpha$ (la stessa in ogni punto). La fase dunque dipende, oltre che da $\alpha$, anche dal parametro $C$, che rappresenta ogni possibile curva chiusa passante per $P$. Il cambiamento totale di fase intorno ad una di queste curve chiuse è dato dalla circuitazione del gradiente della fase intorno alla curva, a cui dà contributo il solo termine polidromo $\beta$: $\Delta\varphi = \oint_C \nabla\varphi \cdot ds = \oint_C \nabla\gamma \cdot ds + \oint_C \nabla\beta \cdot ds + \oint_C \nabla\alpha \cdot ds = \oint_C \nabla\beta \cdot ds = \beta$ [14].

Questo risultato dimostra che il cambiamento totale di fase intorno a una curva chiusa è uguale all'indeterminazione $\beta$ della fase in ogni suo punto, assunto come punto di partenza e d'arrivo del percorso d'integrazione intorno alla curva.

Mediante $\beta$ si può scrivere $\psi$ nella forma $\psi = \psi_1 \cdot e^{i\beta}$, dove $\psi_1$ è una funzione d'onda di modulo uguale al modulo di $\psi$ e con fase univocamente determinata (a meno della costante ininfluente già vista) in ogni punto[15]. L'indeterminazione della fase di $\psi$ è interamente contenuta nel fattore $e^{i\beta}$. La fase (univoca) di $\psi_1$ non dà contributo al cambiamento totale di fase intorno a una curva chiusa. Una fase con cambiamento totale non nullo intorno a una curva chiusa si dice non integrabile.

Affinché la non integrabilità della fase, connessa al fattore di fase indeterminato nella funzione d'onda, sia compatibile con la teoria quantistica è necessario che non insorgano ambiguità in nessuna applicazione della teoria, ossia che la teoria resti consistente. Si può dimostrare che la consistenza matematica è fatta salva sotto determinate condizioni.

*Una questione di coerenza*

Che cosa intendiamo, esattamente, quando parliamo di consistenza matematica della teoria quantistica, e quali sono le condizioni da imporre per salvaguardarla? Perché la teoria resti consistente in presenza di fasi non integrabili nelle funzioni d'onda, è necessario che tutto ciò che nella teoria quantistica ha immediato significato fisico sia determinato univocamente. In tutte le applicazioni della teoria l'indeterminazione della fase deve essere compensata, nel corso delle operazioni intermedie, da un'indeterminazione di segno opposto, sufficiente a rendere il risultato finale, che esprime il contenuto fisico, perfettamente determinato. Qualsiasi risultato con un significato fisico diretto deve avere pertanto un cambiamento totale di fase nullo ($\beta = 0$) lungo un percorso chiuso. Naturalmente, se il risultato finale è una funzione d'onda - che generalmente non ha un significato fisico immediato - questa deve conservare la necessaria indeterminazione.

Nelle applicazioni più semplici della meccanica quantistica la funzione d'onda $\psi$ viene moltiplicata per la sua complessa coniugata $\phi = \psi^*$, ottenendo una "funzione densità" $\phi\psi$, che ha un immediato significato fisico[16]. Poiché le fasi di una funzione d'onda e della sua complessa coniugata sono uguali ed opposte, esse si cancellano nel prodotto ($-\varphi + \varphi \equiv 0$ e, *a fortiori*, $\beta \equiv 0$), e di conseguenza la funzione densità è sempre indipendente dalla fase. Va da sé, quindi, che l'indeterminazione della fase della funzione d'onda non causa nessuna ambiguità della funzione densità.

In operazioni più generali interviene il prodotto $\phi_m \psi_n$[17] di due diverse funzioni d'onda $\psi_m$ e $\psi_n$. In questo caso l'indeterminazione della fase è $\beta = -\beta_m + \beta_n$ più la costante ininfluente già vista (v. nota [12]). La condizione che il cambiamento di fase sia nullo intorno a una curva chiusa $\beta = -\beta_m + \beta_n = 0$ implica che i cambiamenti di fase di $\phi_m$ e $\psi_n$ intorno a una curva chiusa siano uguali e opposti, e quindi siano gli stessi per le due funzioni d'onda: $\beta_m = \beta_n$.

Dirac ottiene in tal modo il risultato generale che «*il cambiamento di fase di una funzione d'onda intorno a una qualsiasi curva chiusa deve essere lo stesso per tutte le funzioni d'onda*».

Il ragionamento di Dirac prosegue così:
«Questa condizione, quando viene estesa per dare la stessa incertezza di fase per le funzioni di trasformazione e le matrici che rappresentano osservabili (con riferimento a rappresentazioni diagonali in *x*, *y* e *z*) così come per le funzioni d'onda, è sufficiente ad assicurare che la non integrabilità della fase non dà origine ad ambiguità in tutte le applicazioni della teoria. Ovunque appare una $\psi_n$, se non è moltiplicata per una $\phi_m$, sarà in ogni caso moltiplicata per qualcosa di una natura simile a $\phi_m$, ottenendo una cancellazione dell'incertezza di fase, eccetto per una costante ininfluente. Per esempio, se $\psi_n$ deve essere trasformata in un'altra rappresentazione in cui, diciamo, le osservabili $\xi$ sono diagonali, deve essere moltiplicata per la funzione di trasformazione ($\xi$,*xyzt*) e integrata rispetto a *x*, *y* e *z*. Questa funzione di trasformazione avrà la stessa incertezza di fase di una $\phi$, così che la funzione d'onda trasformata avrà la fase determinata, eccetto per una costante indipendente da $\xi$[18]. Ancora, se moltiplichiamo $\psi_n$ per una matrice $\langle x'y'z't|\hat{\alpha}|x''y''z''t\rangle$, che rappresenta un'osservabile $\alpha$, l'incertezza di fase sulla colonna [specificata da $x'',y'',z'',t$] cancellerà l'incertezza su $\psi_n$ e l'incertezza sulla riga sopravvivrà e fornirà la necessaria incertezza nella nuova funzione d'onda $\alpha\psi_n$[19]. Il principio di sovrapposizione per le funzioni d'onda verrà discusso un pò più avanti e quando questo punto verrà stabilito completerà la dimostrazione che tutte le operazioni della meccanica quantistica possono esser portate a compimento esattamente come se non vi fosse nessuna incertezza di fase. Il precedente risultato che il cambiamento di fase attorno a una curva chiusa deve essere lo stesso per tutte le funzioni d'onda significa che questo cambiamento di fase deve esser determinato dal sistema dinamico stesso (e forse in parte dalla rappresentazione) e deve essere indipendente dallo stato del sistema considerato. Poiché il nostro sistema dinamico è meramente una semplice particella, sembra che la non integrabilità della fase debba esser connessa col campo di forza in cui si muove la particella».

Matematicamente la non integrabilità della fase richiede che $\beta$ non abbia un valore definito in ciascun punto particolare, ma le sue derivate $k = \nabla \beta = (\frac{\partial \beta}{\partial t}, \frac{\partial \beta}{\partial x}, \frac{\partial \beta}{\partial y}, \frac{\partial \beta}{\partial z})$ sono ivi definite, sebbene non soddisfino in generale le condizioni d'integrabilità $\partial k_x / \partial y = \partial k_y / \partial x$, ecc. (che si possono esprimere globalmente come $rot\, k \neq 0$). Il cambiamento totale di fase intorno a una curva chiusa $\beta = \oint k \cdot ds$, che è lo stesso per tutte le funzioni d'onda, sarà allora, applicando il teorema di Stokes generalizzato a 4 dimensioni,

$$\beta = \oint_C k \cdot ds = \oiint_S rot\, k \cdot dS = \oiint_S rot\, k \cdot dr \wedge dr' = \oiint_S (\mathbf{grad}\, k_0 - \frac{\partial \mathbf{k}}{\partial t}, rot\, \mathbf{k}) \cdot (\mathbf{dS}^0, \mathbf{dS}),$$

dove *S* è una qualsiasi superficie (bidimensionale immersa in *4* dimensioni) delimitata dalla curva chiusa *C*, *k* e *ds* (elemento di linea della curva chiusa) sono quadrivettori (*4*-vettori), $rot\, k = (\mathbf{grad}\, k_0 - \frac{\partial \mathbf{k}}{\partial t}, rot\, \mathbf{k})$ e $dS = dr \wedge dr' = (\mathbf{dS}^0, \mathbf{dS}) = (\mathbf{dr}dt' - \mathbf{dr}'dt, \mathbf{dr} \wedge \mathbf{dr}')$ (elemento della superficie bidimensionale) sono esavettori (*6*-vettori).

Si dimostra facilmente l'unicità di *rotk* per tutte le superfici delimitate dalla stessa curva chiusa: allora *rotk* è lo stesso per tutte le funzioni d'onda e *k* è definito per ogni funzione d'onda a meno del gradiente di uno scalare[20].

Si può quindi esprimere l'indeterminazione della fase di una funzione d'onda come

$$\beta = \oint (k + \mathrm{grad}\chi) \cdot ds = \oint k \cdot ds$$

con lo stesso *k* per tutte le funzioni d'onda. Di conseguenza, prese due funzioni d'onda $\psi_m = \psi_{1m} e^{i\beta_m}, \psi_n = \psi_{1n} e^{i\beta_n}$, per qualsiasi loro combinazione lineare si ottiene $c_m \psi_m + c_n \psi_n = c_m \psi_{1m} e^{i\beta_m} + c_n \psi_{1n} e^{i\beta_n} = c_m \psi_{1m} e^{i\oint k \cdot ds} + c_n \psi_{1n} e^{i\oint (k + \mathrm{grad}\chi) \cdot ds} = (c_m \psi_{1m} + c_n \psi_{1n}) e^{i\beta}$, dove abbiamo espresso con lo stesso *k* l'indeterminazione di fase delle funzioni d'onda, che è uguale alla differenza di fase intorno a una possibile curva d'integrazione chiusa, ed è la stessa per tutte le funzioni d'onda (ma possiamo richiedere che *k* differisca per il gradiente di una funzione scalare per diverse funzioni d'onda, senza nessuna differenza nelle nostre conclusioni). Perciò la funzione d'onda risultante da una qualsiasi combinazione lineare di funzioni d'onda, tutte col medesimo cambiamento di fase intorno a una curva chiusa, ha anch'essa lo stesso cambiamento di fase intorno alla stessa curva. Questo consente di dimostrare che la non integrabilità della fase è consistente col principio di sovrapposizione, e completa la dimostrazione che «tutte le operazioni della meccanica quantistica possono esser portate a compimento esattamente come se non vi fosse incertezza di fase alcuna».

### *La connessione fase-campo*

È giunto ora il momento di discutere il significato fisico di questi risultati. Nella nota [15] abbiamo fornito la dimostrazione che se la funzione d'onda $\psi$ soddisfa un'equazione d'onda in cui figurano gli operatori quantità di moto $\hat{\mathbf{p}}$ ed energia $\hat{W}$, $\psi_1$ soddisferà la corrispondente equazione ottenuta sostituendo nella prima gli operatori $\hat{\mathbf{p}}_1 = \hat{\mathbf{p}} + \hbar \mathbf{k}$ e $\hat{W}_1 = \hat{W} - \hbar k_0$ rispettivamente. Era già noto, prima che Dirac avanzasse la teoria del monopolo che, se una particella libera, in assenza di campo elettromagnetico (per esempio un elettrone di carica –*e*, ma lo stesso vale per una generica particella di carica *q*), viene descritta da una funzione d'onda $\psi$, la stessa particella in un campo elettromagnetico di potenziale $A = (A_0, \mathbf{A})$ può esser descritta da una funzione d'onda $\psi_1$, legata a $\psi$ attraverso il fattore di fase $e^{i\beta}$ dalla relazione $\psi = \psi_1 \cdot e^{i\beta}$, dove $\beta = \oint k \cdot ds$ e $k = (k_0, \mathbf{k})$ con $k_0 = -\frac{q}{\hbar} A_0, \quad \mathbf{k} = \frac{q}{\hbar c} \mathbf{A}$.

Poiché $\psi_1$ è una funzione d'onda ordinaria con una fase perfettamente definita in ogni punto, *il campo elettromagnetico si manifesta nella teoria di Dirac come l'effetto di un fattore di fase non integrabile e, reciprocamente, l'effetto del campo rende la fase non integrabile.*

In altri termini, $\psi$ e $\psi_1$ forniscono due descrizioni equivalenti: $\psi_1$ descrive una particella con una fase definita in un campo elettromagnetico; $\psi$ descrive la stessa particella libera, in assenza di campo, con una fase che cambia percorrendo una curva chiusa.

Ricordando le equazioni che legano i potenziali ai campi:

$$\mathbf{H} = \mathrm{rot}\mathbf{A}, \quad \mathbf{E} = -\mathrm{grad}A_0 - \frac{1}{c}\frac{\partial \mathbf{A}}{\partial t},$$

otteniamo:

$$\mathrm{rot}\mathbf{k} = \frac{q}{\hbar c}\mathbf{H}, \quad \mathrm{grad}k_0 - \frac{\partial \mathbf{k}}{\partial t} = \frac{q}{\hbar}\mathbf{E},$$

ed infine:

$$\beta = \oiint_S (\mathbf{grad}\,k_0 - \frac{\partial \mathbf{k}}{\partial t}, rot\mathbf{k}) \cdot (\mathbf{dS}^0, \mathbf{dS}) = \frac{q}{\hbar c} \oiint_S (\mathbf{E}, \mathbf{H}) \cdot (c\mathbf{dS}^0, \mathbf{dS})\,,$$

dove compare il *6-vettore* campo elettromagnetico (**E**,**H**). L'ultima uguaglianza dice che il cambiamento di fase $\beta$ intorno a una curva chiusa è uguale al flusso elettromagnetico uscente da una qualsiasi superficie bidimensionale *S*, immersa nello spazio a quattro dimensioni delimitata dalla curva.

Fino a questo punto, in realtà, Dirac non aveva ottenuto nessun risultato sostanzialmente nuovo. Il suo vero ed originale contributo fu una nuova interpretazione fisica di risultati noti.

Se consideriamo separatamente le componenti elettriche e magnetiche del *6-vettore* (**E**,**H**) troviamo che il cambiamento di fase intorno a una curva chiusa, connesso al campo elettrico $\beta_{el} = \frac{q}{\hbar}\oiint_S \mathbf{E}\cdot\mathbf{dS}^0 = \oint k_0 dt$ è nullo quando i potenziali non dipendono esplicitamente dal tempo[21]. Pertanto in condizioni stazionarie il cambiamento totale di fase intorno a una curva chiusa è connesso al solo campo magnetico, ed è legato al flusso magnetico uscente da una superficie *S* (che questa volta è una superficie bidimensionale immersa nello spazio tridimensionale ordinario) che ha per contorno la proiezione nello spazio tridimensionale della curva chiusa considerata. Si ottiene allora la seguente relazione fase-campo magnetico:

$$\beta = \beta_{magn} = \oint \mathbf{k}\cdot\mathbf{ds} = \frac{q}{\hbar c}\oiint_S \mathbf{H}\cdot\mathbf{dS} = \frac{q}{\hbar c}\Phi_H\,.$$

Se ora consideriamo una superficie a forma di sacco, e la chiudiamo rimpicciolendo via via la curva che la delimita, il cambiamento di fase si riduce a zero con lo svanire della curva chiusa. In questa situazione si annulla anche il flusso magnetico $\Phi_H$ che attraversa la superficie (e qualsiasi altra superficie chiusa ottenuta nello stesso modo), e si ritrova l'ordinaria legge dell'assenza di sorgenti del campo magnetico di cui abbiamo parlato all'inizio.

Come abbiamo detto, la vera novità introdotta da Dirac fino a questo punto del ragionamento fu il concetto che la non integrabilità della fase è connessa al campo elettromagnetico. Questa nuova interpretazione condusse ad una migliore comprensione della natura dell'interazione elettromagnetica, e mostrò che si potevano ottenere nuove acquisizioni approfondendo l'indagine matematica dei fondamenti della meccanica quantistica. In particolare dimostrò che le fasi non integrabili sono possibili nella meccanica quantistica introducendo un campo elettromagnetico, che ne fornisce l'interpretazione fisica: il campo elettromagnetico è connesso con una fase non integrabile, e viceversa. In precedenza questa connessione era stata messa in evidenza da Weyl in un importante saggio[22], dove per la prima volta veniva introdotto il principio d'invarianza di *gauge* nella teoria quantistica. Altri ricercatori avevano raggiunto acquisizioni simili affrontando il problema della connessione fase-campo da differenti punti di vista. Dirac invece ragionò sulle proprietà matematiche generali delle funzioni d'onda e sulla condizione di non integrabilità della fase. Con soddisfazione, poté concludere che «fasi non integrabili sono perfettamente compatibili con tutti i principi generali della meccanica quantistica e non ne restringono in alcuna misura l'interpretazione fisica».

### *Hic sunt leones: poli, stringhe, singolarità*

Abbiamo visto finora che le derivate non integrabili *k* della fase della funzione d'onda vengono interpretate nei termini dei potenziali elettromagnetici. Con questa assunzione la teoria di Dirac diviene equivalente all'usuale teoria del moto di una carica in un campo elettromagnetico e non introduce alcuna legge nuova. Viene invece introdotta una nuova interpretazione fisica: la non integrabilità della fase è una manifestazione *gauge*-invariante del campo elettromagnetico.

Un fisico "normale" si sarebbe accontentato di questo risultato, ma non Dirac! L'interpretazione appena proposta si rivela subito insufficiente allorché si tiene conto di un nuovo elemento finora non considerato: la fase, come ogni angolo, è sempre determinata a meno di un arbitrario multiplo di $2\pi$[23]. Le due funzioni d'onda $\psi$ e $\psi e^{i2\pi n}$ rappresentano dunque lo stesso identico stato fisico. Questo fatto implica una riconsiderazione della connessione fase-campo e produce importanti conseguenze fisiche. Infatti il cambiamento di fase intorno a una curva chiusa può ora differire per diverse funzioni d'onda di multipli arbitrari di $2\pi$ (rispettivamente, diciamo $\beta$ per $\psi$ e $\beta + 2\pi n$ per $\psi e^{i2\pi n}$). Di conseguenza, il cambiamento di fase non può trovare immediata interpretazione, come prima, nei termini dei potenziali elettromagnetici.

Seguiamo ancora una volta il ragionamento di Dirac. «Consideriamo dapprima una curva chiusa molto piccola. L'equazione d'onda richiede la continuità della funzione d'onda[24] (tranne in circostanze molto particolari che possiamo trascurare nel nostro contesto) e quindi il cambiamento di fase intorno ad una piccola curva chiusa deve essere piccolo. Allora questo cambiamento [di fase] non può differire per multipli di $2\pi$ per diverse funzioni d'onda. Deve avere un valore definito e si può perciò interpretare senza ambiguità nei termini del flusso del *6*-vettore (**E**,**H**) attraverso la piccola curva chiusa, ed anche il flusso deve essere piccolo. Vi è tuttavia un'eccezione quando la funzione d'onda si annulla, poiché in questo caso la sua fase è priva di significato[25]. Essendo una funzione complessa, il suo annullarsi richiederà due condizioni, sì che in generale i punti in cui si annulla giaceranno lungo un linea[26]. Chiamiamo *linea nodale* una tale linea». La congettura che una funzione d'onda debba annullarsi lungo una linea (o superficie) nodale è nota come "veto di Dirac" e la linea (superficie) nodale è proprio la "stringa di Dirac" che abbiamo incontrato parlando del semi-solenoide infinito. Tale congettura equivale ad imporre che una particella abbia una probabilità uguale a zero di trovarsi sulla stringa di Dirac, ovvero che non possa mai attraversarla. La stringa di Dirac sarebbe perciò del tutto inaccessibile. Evidentemente nulla impone che una funzione d'onda debba avere una linea (o superficie) nodale, ma nulla lo vieta in linea di principio: se per qualche ragione funzione d'onda non potesse annullarsi (se non all'infinito) cadrebbe il veto di Dirac, e con esso verrebbe invalidata l'intera teoria del monopolo.

«Se ora – prosegue Dirac - prendiamo una funzione d'onda con una linea nodale che passa attraverso la nostra piccola curva chiusa, le considerazioni di continuità non ci consentiranno più di affermare che il cambiamento di fase intorno alla piccola curva chiusa deve essere piccolo. Tutto ciò che potremo dire è che il cambiamento di fase deve essere prossimo a $2\pi n$ dove $n$ è un intero, positivo o negativo[27]. Questo intero [$n$] sarà una caratteristica della linea nodale. Il suo segno sarà associato con una direzione intorno alla linea nodale, che a sua volta può essere associato con una direzione lungo la linea nodale. La differenza tra il cambiamento di fase intorno alla piccola curva chiusa e il $2\pi n$ più prossimo deve essere lo stesso del cambiamento di fase intorno alla curva chiusa per una funzione d'onda senza linea nodale che l'attraversa[28]. È perciò questa differenza che deve essere interpretata nei termini del flusso del 6-vettore (**E**,**H**) attraverso la curva chiusa». Quindi la differenza di fase è data in generale da:

$$\beta = 2\pi n + \oiint_S \frac{q}{\hbar c}(\mathbf{E},\mathbf{H}) \cdot (c\mathbf{dS}^0, \mathbf{dS}),$$

e per una curva chiusa nello spazio tridimensionale, in cui interviene il solo flusso magnetico, si ottiene:

$$\beta_{\text{magn}} = 2\pi n + \oiint_S \frac{q}{\hbar c}\mathbf{H} \cdot \mathbf{dS} = 2\pi n + \frac{q}{\hbar c}\Phi_H .$$

Se si rinunciasse alla congettura del "veto di Dirac" non si potrebbe avere la singolarità del potenziale vettore, che è necessaria per ottenere la distribuzione isotropa che caratterizza le linee di forza magnetiche del polo isolato (v. *Fig. 2*).[29]

Si può trattare lo stesso problema per una curva chiusa grande, dividendola in un reticolo di piccole curve chiuse che giacciono su una superficie $S$ delimitata dalla grande curva chiusa $s$. Il cambiamento totale di fase intorno alla grande curva sarà allora uguale alla somma di tutti i cambiamenti intorno alle piccole curve, cioè sarà:

$$\beta = 2\pi \sum n + \frac{q}{\hbar c} \cdot \Phi_H ,$$

dove il flusso viene preso su tutta la superficie $S$, e la somma viene calcolata su tutte le linee nodali che la attraversano, con il rispettivo segno per ogni termine.

In questa espressione la parte $\frac{q}{\hbar c} \cdot \Phi_H$ deve essere la stessa per tutte le funzioni d'onda, mentre la parte $2\pi \sum n$ può esser diversa per diverse funzioni d'onda. L'espressione fornisce, per ogni superficie $S$, il cambiamento di fase intorno al contorno $s$. Se la superficie è chiusa non ha contorno, e il cambiamento di fase si deve annullare (per tutte le funzioni d'onda). Di conseguenza $\sum n$ sommata su tutte le linee nodali che attraversano una superficie chiusa deve essere la stessa per tutte le funzioni d'onda, e deve essere uguale a $-\frac{q}{2\pi \hbar c} \Phi_H$. Se $\sum n$ non si annulla significa che alcune linee nodali devono avere dei punti estremi all'interno della superficie chiusa, perché altrimenti una linea nodale deve attraversare la superficie almeno due volte e fornisce alla sommatoria due contributi uguali ed opposti nei due punti dove "perfora" la superficie. Il valore di $\sum n$ per una superficie chiusa sarà allora uguale alla somma algebrica dei valori di $n$ per tutte le linee nodali con punti estremi all'interno della superficie, e deve essere lo stesso per tutte le funzioni d'onda. Poiché questa condizione vale per qualsiasi superficie chiusa, ne consegue che «i punti estremi devono essere gli stessi per tutte le funzioni d'onda. Questi punti estremi sono allora punti di singolarità nel campo elettromagnetico». Una singolarità nel campo elettromagnetico è caratterizzata dal fatto di irradiare un flusso netto attraverso una qualsiasi superficie chiusa che la circonda. Il flusso magnetico totale attraverso una *piccola* superficie chiusa che circonda uno di questi punti è $4\pi g = 2\pi n \hbar c / q$ [30], dove $n$ è la caratteristica della linea nodale con un punto estremo all'interno della superficie, o la somma algebrica delle caratteristiche di tutte le linee nodali con punti estremi all'interno, quando ve n'è più d'una. Quindi nel punto estremo vi sarà un polo magnetico d'intensità $g = \frac{1}{2} n \hbar c / q$, che irradia all'esterno il flusso quantizzato $2\pi n \hbar c / q$, trasportato nel monopolo attraverso la stringa di Dirac. All'altro estremo, che si trova all'infinito, vi sarà un polo di polarità opposta (antimonopolo), che fornisce il flusso necessario a mantenere il bilancio in parità.

La teoria permette dunque poli isolati, ma questi poli devono avere una carica magnetica quantizzata, con un quanto fondamentale $g_0$ legato alla carica elementare $e$ dalla condizione (di quantizzazione) $e g_0 = \frac{1}{2} \hbar c$, che possiamo esprimere nei termini della costante d'accoppiamento (costante di struttura fine) $\frac{e^2}{\hbar c} \approx \frac{1}{137}$, come $g_0 = \frac{1}{2} \frac{\hbar c}{e^2} e \approx \frac{1}{2} 137 e = 68{,}5 e$.

Reciprocamente, anche la carica elettrica deve essere quantizzata con la condizione $q = \frac{1}{2} n \hbar c / g$,

ovvero $e = \frac{1}{2} \hbar c / g_0 = 2 \frac{e^2}{\hbar c} g_0 \approx 2 \frac{1}{137} g_0$ per la carica elementare.

Scrivendo le condizioni precedenti come $qg = \frac{1}{2} n\hbar c$, che equivale alla (1), nel caso generale e $eg_0 = \frac{1}{2}\hbar c$ per le cariche elementari, si ottiene $g = ng_0$ (per $q = e$) e, reciprocamente, $q = ne$ (per $g = g_0$).

Quest'ultima condizione significa che la carica di una particella in moto intorno a un monopolo deve essere un multiplo intero della carica elementare affinché la sua funzione d'onda possa esistere. Se venisse identificato un monopolo di carica magnetica, poniamo, $g_0 \approx \frac{137}{2} \cdot \frac{1}{3} e$, allora la carica elementare sarebbe $\frac{1}{3} e$, cioè esisterebbero cariche frazionarie (in unità di carica dell'elettrone).

Se mettiamo a confronto le costanti d'accoppiamento elettrica $\frac{e^2}{\hbar c} \approx \frac{1}{137} \ll 1$ e magnetica $\frac{g_0^2}{\hbar c} = \frac{1}{4}\left(\frac{\hbar c}{e^2}\right) \approx \frac{1}{4} 137 \approx 34 \gg 1$ notiamo un'importante differenza quantitativa: le interazioni tra cariche elettriche sono molto più deboli di quelle tra cariche magnetiche, e sono trattabili con sviluppi perturbativi fino al grado di precisione desiderato; le interazioni tra cariche magnetiche sono molto più difficili da affrontare poiché non si può far uso di tali metodi.

## *Una lezione di Fermi*

Si può ricavare la condizione di quantizzazione (1) anche in altri modi. Uno dei più semplici fu suggerito da Fermi nel 1950.[31]

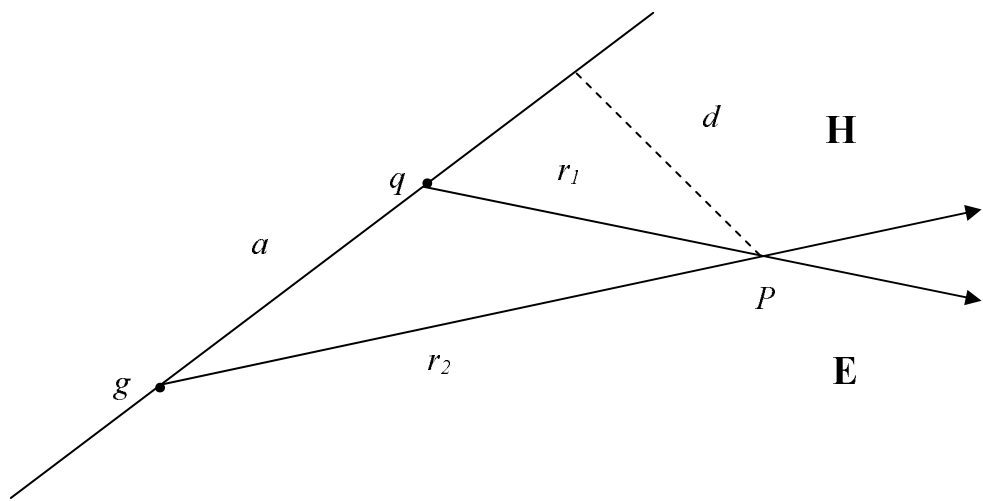

*Fig. 4*

Fermi calcolò il momento angolare del campo elettromagnetico rispetto all'asse $a$ (v. *Fig. 4*). Per il campo elettromagnetico nel punto $P$ abbiamo: $\mathbf{E} = \frac{q}{r_1^2}\hat{\mathbf{r}}_1$, $\mathbf{H} = \frac{g}{r_2^2}\hat{\mathbf{r}}_2$.

Scegliamo di lavorare soltanto sugli ordini di grandezza e poniamo $r_1 \approx r_2 \approx d \approx a$. Possiamo allora scrivere per il campo elettromagnetico $\mathbf{E} \approx \dfrac{q}{a^2}\hat{\mathbf{r}}_1$, $\mathbf{H} \approx \dfrac{g}{a^2}\hat{\mathbf{r}}_2$, e per la sua quantità di moto $\dfrac{1}{4\pi c}|\mathbf{E} \wedge \mathbf{H}| \approx \dfrac{qg}{a^4 c}$.

Il momento angolare si ottiene moltiplicando questa quantità per $d$ ed integrando su un volume delle dimensioni di $a^3$. Pertanto, essendo $d \approx a$, il momento angolare risultante sarà $\approx a \cdot \left(\dfrac{qg}{a^4 c}\right) \cdot a^3 = \dfrac{qg}{c}$.

Allora per la quantizzazione del momento angolare si ottiene $\dfrac{qg}{c} = n\dfrac{\hbar}{2}$, in accordo col risultato di Dirac.

### *Un volo in mongolfiera*

Le prime ricerche sperimentali di monopoli si concentrarono sullo studio dei raggi cosmici, impiegando lastre sensibili alle radiazioni ionizzanti a bordo di palloni lanciati ad alta quota e poi recuperati.

I raggi cosmici sono una radiazione di origine extraterrestre, di composizione molto complessa, particolarmente ricchi di protoni d'alta energia. Nell'interazione dei protoni cosmici con l'atmosfera si verifica una grande varietà di fenomeni, come la produzione di sciami di particelle pesanti (p.es. iperoni) che interagiscono a loro volta con l'atmosfera, dando origine ad ulteriori tipi di particelle e radiazioni. Questo complesso di fenomeni si estende su uno spettro energetico molto ampio comprendente una lunga serie di processi nucleari che si verificano nell'alta atmosfera terrestre e nello spazio. La composizione nucleare della radiazione cosmica va dai nuclei leggeri, come $^4$He e $^6$Li, a nuclei pesanti come quelli di Erbio (Z = 68), che come vedremo più avanti possono venire facilmente confusi coi monopoli.

Le tracce dei nuclei pesanti nelle lastre dei rivelatori presentano un aspetto molto caratteristico. Le tracce dei protoni sono sottili e diritte, mentre quelle dei nuclei pesanti sono larghe, marcate, fortemente asimmetriche, e vanno riducendosi di spessore lungo il percorso, sfumando via via. Le tracce di un monopolo somiglierebbero a quelle di un nucleo pesante, ma sarebbero molto più simmetriche. Accade però sovente che le tracce dei nuclei pesanti appaiano confuse e non possano venir usate per un confronto. Tuttavia negli anni '50 furono osservate tracce molto marcate e simmetriche, che facevano pensare a monopoli, ma questa ipotesi non trovò conferma. Un evento notevole venne osservato nel 1975 in una lastra che aveva viaggiato su un pallone sonda. In una comunicazione al «Physical Review Letters»[32], che ebbe un notevole eco anche sulla stampa non specializzata, venne annunciata la possibile scoperta del primo monopolo magnetico, ma anche in questo caso prevalse l'interpretazione, sostenuta autorevolmente da L. W. Alvarez, che si trattasse in realtà di un nucleo pesante.

Le ricerche sui raggi cosmici non hanno fornito prove convincenti dell'esistenza di monopoli, almeno finora. Per dare un'idea delle enormi difficoltà di queste ricerche, considereremo il problema del passaggio di un monopolo attraverso la materia, per es. l'aria.

Il problema generale del rallentamento nella materia di una particella carica pesante fu affrontato per la prima volta da Bohr nel 1913[33] in un contesto classico, considerando il passaggio di una particella pesante di carica $Z_1 e$ in moto con velocità $v$ nelle vicinanze di un elettrone di carica $e$ e massa $m$, a una distanza minima (parametro d'impatto) $b$ (v. *Fig. 5*).

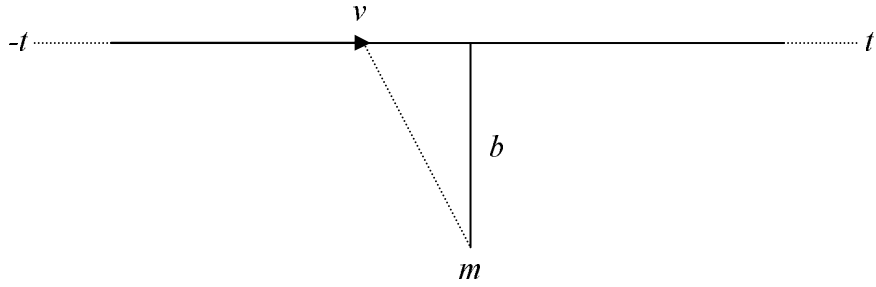

*Fig. 5*

Si assume l'ipotesi semplificatrice che la particella, supposta molto più pesante dell'elettrone, non subisca praticamente nessuna deflessione dall'iniziale traiettoria rettilinea. Si assume inoltre che l'elettrone sia libero e inizialmente in quiete, e nella collisione acquisti una velocità $\Delta v_e \ll v$. Si ammette infine che durante la collisione l'elettrone si muova così poco dalla sua posizione iniziale (ossia si muova al massimo di $\Delta b \ll b$) che nel calcolo del campo elettrico agente su di esso e dovuto alla particella pesante si possa assumere $b$ costante[34]. Con queste ipotesi si ottiene per il campo elettrico trasversale agente sull'elettrone:

$$E_\perp(t) = \frac{Z_1 e}{b^2 + v^2 t^2} \cdot \frac{b}{\sqrt{b^2 + v^2 t^2}} = \frac{Z_1 e}{b^2} \cdot \frac{1}{\left[1+\left(vt/b\right)^2\right]^{\frac{3}{2}}} = E_{\max}^{(Z_1)} \cdot \frac{1}{\left[1+\left(vt/b\right)^2\right]^{\frac{3}{2}}},$$

dove $E_{\max}^{(Z_1)} = \frac{Z_1 e}{b^2}$; per il campo longitudinale: $E_\parallel(t) = E_{\max} \cdot \dfrac{vt/b}{\left[1+\left(vt/b\right)^2\right]^{\frac{3}{2}}}$; e per l'impulso (trasversale[35]) conferito all'elettrone:

$$\Delta p = \Delta p_\perp = \int_{-\infty}^{+\infty} e E_\perp(t) dt = e E_{\max}^{(Z_1)} \cdot \int_{-\infty}^{+\infty} \frac{1}{\left[1+\left(vt/b\right)^2\right]^{\frac{3}{2}}} dt = e E_{\max}^{(Z_1)} \cdot \left(\frac{b}{v}\right) \cdot \int_{-\infty}^{+\infty} \frac{1}{(1+\xi^2)^{\frac{3}{2}}} d\xi = e E_{\max}^{(Z_1)} \cdot \left(\frac{b}{v}\right) \cdot \left[\frac{\xi}{\sqrt{1+\xi^2}}\right]_{-\infty}^{+\infty} = e E_{\max}^{(Z_1)} \cdot \left(\frac{2b}{v}\right)$$

($\xi = \frac{vt}{b}$). L'energia trasferita all'elettrone dalla particella carica è pertanto:

$$\Delta W_{Z_1} = \frac{(\Delta p)^2}{2m} = \frac{(\Delta p_\perp)^2}{2m} = e^2 E_{\max}^{(Z_1)2} \cdot \left(\frac{2b^2}{mv^2}\right).$$

Si può fare un calcolo analogo per il rallentamento di un monopolo (che si assume sia molto più pesante di un elettrone) di carica magnetica $g$ nel suo passaggio attraverso la materia. In questo caso il campo elettrico agente sull'elettrone è legato al potenziale vettore $\mathbf{A}$ dall'equazione $\mathbf{E} = -\dfrac{1}{c}\dfrac{\partial \mathbf{A}}{\partial t}$, dove $\mathbf{A} = \dfrac{g}{r}\dfrac{(1-\cos\theta)}{\sin\theta}\hat{\varphi} = \dfrac{g}{b}\left(1 - \dfrac{vt}{\sqrt{b^2 + v^2 t^2}}\right)\hat{\varphi}$ (con le stesse convenzioni della *Fig. 1*).

Otteniamo per il campo elettrico:

$$\mathbf{E} = -\frac{1}{c}\frac{d\mathbf{A}}{dt} = \frac{v}{c}\cdot\left(\frac{g}{b}\right)\cdot\frac{d}{dt}\left(\frac{t}{\sqrt{b^2+v^2t^2}}\right)\hat{\varphi} = \frac{v}{c}\cdot\left(\frac{g}{b^2}\right)\cdot\left(\frac{b}{v}\right)\cdot\frac{d}{dt}\left(\frac{vt/b}{\sqrt{1+(vt/b)^2}}\right)\hat{\varphi} = E_{\max}^{(g)}\cdot\left(\frac{b}{v}\right)\cdot\frac{d}{dt}\left(\frac{\xi}{\sqrt{1+\xi^2}}\right)\hat{\varphi},$$

dove $E_{\max}^{(g)} = \frac{v}{c}H_{\max}^{(g)} = \frac{v}{c}\left(\frac{g}{b^2}\right)$;

per l'impulso trasferito:

$$\Delta\mathbf{p} = \int_{-\infty}^{+\infty} e\mathbf{E}\,dt = eE_{\max}^{(g)}\cdot\left(\frac{b}{v}\right)\cdot\int_{-\infty}^{+\infty}\frac{d}{dt}\left(\frac{\xi}{\sqrt{1+\xi^2}}\right)dt\,\hat{\varphi} = eE_{\max}^{(g)}\cdot\left(\frac{b}{v}\right)\cdot\left[\frac{\xi}{\sqrt{1+\xi^2}}\right]_{-\infty}^{+\infty}\hat{\varphi} = eE_{\max}^{(g)}\cdot\left(\frac{2b}{v}\right)\hat{\varphi},$$

e per il trasferimento d'energia: $\Delta W_g = \frac{(\Delta p)^2}{2m} = e^2 E_{\max}^{(g)\,2}\cdot\left(\frac{2b^2}{mv^2}\right)$.

La trattazione del moto di un monopolo attraverso la materia è pertanto equivalente a quella del moto di una carica pesante con la sostituzione di $E_{\max}^{(Z_1)}$ con $E_{\max}^{(g)}$.

Il rapporto tra le energie trasferite (connesse alle ionizzazioni rispettivamente prodotte nel mezzo attraversato) è pertanto:

$$R = \frac{\Delta W_g}{\Delta W_{Z_1}} = \frac{E_{\max}^{(g)\,2}}{E_{\max}^{(Z_1)\,2}} = \frac{\left(\frac{v}{c}\right)^2\cdot\left(\frac{g}{b^2}\right)^2}{\left(\frac{Z_1 e}{b^2}\right)^2} = \left(\frac{v}{c}\right)^2\cdot\left(\frac{g}{Z_1 e}\right)^2.$$

Se $v \cong c$, cioè per monopoli in moto con velocità relativistiche $R \cong \left(\frac{g}{Z_1 e}\right)^2 \simeq \left(\frac{68,5}{Z_1}\right)^2 \approx \frac{4700}{Z_1^2}$.

Ponendo $R \cong 1$ si vede che un monopolo relativistico produrrebbe nella materia una ionizzazione equivalente a quella di un nucleo atomico pesante di carica $Z_1 \simeq 68$ (Erbio). Ciò spiega perché le tracce delle due particelle in una lastra sensibile possono essere facilmente confuse.

Per calcolare il potere frenante della materia su una particella carica pesante consideriamo il numero di collisioni nell'unità di percorso, con parametro d'urto tra $b$ e $b+db$. Questo numero è pari al numero di elettroni contenuto nella buccia cilindrica $2\pi b\,db$ di spessore $db$ (di lunghezza unitaria). Otteniamo il numero di elettroni per unità di volume $\mathcal{N} = NZ_2$ moltiplicando il numero di atomi per unità di volume $N = N_0\rho/A$ (dove $N_0 = 6,02\cdot 10^{23}$ è il numero di Avogadro, $\rho$ la densità del mezzo attraversato, $A$ il numero di massa del mezzo) per il numero atomico $Z_2$ del mezzo attraversato[36]. Moltiplicando ancora $\mathcal{N}$ per $2\pi b\,db$ otteniamo il numero di elettroni nella buccia $2\pi\mathcal{N}b\,db$, da cui ricaviamo l'energia $\Delta W\cdot 2\pi\mathcal{N}b\,db$ persa nella buccia. Integrando poi su tutti i possibili valori del parametro d'urto otteniamo infine la perdita d'energia nell'unità di percorso $-\frac{d\mathcal{E}}{dx}$, cioè il potere frenante $\mathcal{F}$:

$$\mathcal{F} = -\frac{d\mathcal{E}}{dx} = 2\pi\mathcal{N}\int_{b_{\min}}^{b_{\max}}\Delta W\cdot b\,db = 2\pi\mathcal{N}\frac{e^2}{2m}\int_{b_{\min}}^{b_{\max}}E_{\max}^2(b)\cdot\left(\frac{2b}{v}\right)^2\cdot b\,db = \frac{4\pi\mathcal{N}e^2}{mv^2}\cdot Q^2\int_{b_{\min}}^{b_{\max}}\frac{db}{b} = \frac{4\pi\mathcal{N}e^2}{mv^2}\cdot Q^2\ln\left(\frac{b_{\max}}{b_{\min}}\right),$$

dove $Q = Q_{Z_1} = Z_1 e$ per una particella carica e $Q = Q_g = \left(\frac{v}{c}\right)g$ per un monopolo.

Poiché il fattore logaritmico diverge per $b_{\max}\to\infty$ e $b_{\min}\to 0$ è necessario limitare superiormente e inferiormente i valori di $b_{\max}$ e $b_{\min}$. A questo scopo si introducono il semidiametro di collisione $b_{\min} = \frac{Qe}{mv^2}$ e il raggio adiabatico $b_{\max} = \frac{v}{\omega}$, dove $\omega$ è la frequenza orbitale dell'elettrone[37]. Si ottiene in tal modo la formula di Bohr, nella versione non relativistica[38], della perdita d'energia di una particella carica $\mathscr{F}_{Z_1} = \frac{4\pi \mathscr{N} e^2}{mv^2} \cdot Z_1^2 e^2 \ln\left(\frac{b_{max}}{b_{min}}\right) = \frac{4\pi mc^2 r_e^2 Z_1^2 \mathscr{N}}{\beta^2} \ln\left(\frac{mv^3}{\omega Z_1 e^2}\right)$, dove $mc^2 \approx 0,5 MeV$ è l'energia intrinseca dell'elettrone, $r_e = \frac{e^2}{mc^2} \approx 2,8 \cdot 10^{-13} cm \left(r_e^2 \approx 7,9 \cdot 10^{-26} cm^2\right)$ il raggio classico dell'elettrone e $\beta = \frac{v}{c}$.

Applicando la stessa formula nel caso del monopolo si ottiene:
$$\mathscr{F}_{g_0} = 4\pi mc^2 r_e^2 \mathscr{N} \cdot \left[\frac{1}{2}\frac{\hbar c}{e^2}\right]^2 \ln\left(\frac{mv^2 c}{\omega g_0 e}\right) \approx 4\pi mc^2 r_e^2 \mathscr{N} \cdot 4700 \cdot \ln\left(\frac{2mv^2 c}{\omega \cdot 137 e^2}\right).$$

Tuttavia per i monopoli relativistici la formula precedente non descrive correttamente la perdita d'energia e sono necessarie alcune correzioni relativistiche al termine logaritmico.

L'effetto di tali correzioni (v. nota [38]) è un aumento (relativamente lento) del potere frenante a velocità relativistiche, mitigato da un effetto di densità dovuto alla polarizzazione del mezzo (effetto Fermi)[39]. In assenza di questi effetti, la perdita d'energia delle particelle cariche veloci sarebbe approssimativamente costante a velocità elevate ($\beta \sim 1$) (lento aumento all'aumentare della velocità per l'effetto combinato del fattore logaritmico relativistico e dell'effetto Fermi), mentre sarebbe proporzionale a $\frac{1}{\beta^2}$ (rapido aumento al diminuire della velocità) a velocità inferiori[40]. Questo spiega perché nelle lastre sensibili alle radiazioni ionizzanti le tracce di un nucleo pesante presentano il loro caratteristico aspetto. Per un monopolo l'effetto del fattore $\frac{1}{\beta^2}$ viene esattamente compensato dal fattore $\beta^2$ proveniente dal quadrato di $Q_g = \beta g$, e la sua traccia risulterebbe perciò fortemente simmetrica. Si può esprimere in modo compatto la perdita d'energia di un monopolo in funzione del minimo potere frenante di una particella ionizzante (cioè di una particella con $Z_1 = 1$, p.es. un muone) nel modo seguente: $\mathscr{F}_{g_0} = R \mathscr{F}_{Z_1} = \frac{\beta^2 \left(g_0/e\right)^2}{Z_1^2} \mathscr{F}_{Z_1} \approx \frac{\beta^2}{Z_1^2} \cdot 4700 \cdot \mathscr{F}_{Z_1} = \beta^2 \cdot 4700 \cdot \mathscr{F}_1 \sim 4700 \cdot \mathscr{F}_1$, dove $\mathscr{F}_1$ è il minimo potere frenante, pari a $\simeq 2 MeV/g \cdot cm^{-2}$ [41]. Da questa formula si vede che la perdita d'energia di un monopolo relativistico è circa 4700 volte superiore a quella di una particella di carica unitaria. Questo fatto ha importanti conseguenze. Per esempio, il nostro pianeta ha un campo magnetico che può essere considerato approssimativamente equivalente a quello di un dipolo magnetico di momento $\mu \simeq 7,81 \cdot 10^{15} Tesla \cdot m^3$ collocato al centro della Terra. Accoppiando un monopolo a questo dipolo, l'energia potenziale del monopolo è $E_p^{(g)} \simeq (68,5e)\frac{\mu}{r^2} \cos \vartheta$. Al polo ($\theta = 0$) $E_p^{(g)} \simeq 5 \cdot 10^{12} eV \simeq 5 TeV$ [42], perciò un monopolo acquisterebbe dal campo magnetico terrestre un'energia paragonabile alle massime energie attualmente raggiungibili con le più potenti macchine acceleratici.

Quando un monopolo entra in contatto con l'atmosfera terrestre avviene una ionizzazione e il monopolo ne risulta frenato, con una perdita d'energia di $\sim 4700 \mathscr{F}_1 \sim 10 GeV/g \cdot cm^{-2}$, ovvero $\sim \rho_{(aria)} \cdot 4700 \mathscr{F}_1 \sim 10 MeV/cm$. Tuttavia le linee di forza del campo magnetico terrestre trascinano il monopolo, di modo che il suo moto può risultare accelerato o frenato[43]. In un campo magnetico un monopolo acquista un'energia $\sim g_0 H z$, dove $H$ è il campo magnetico e $z$ la proiezione dello spostamento lungo una linea di forza. L'energia acquistata in un campo magnetico unitario nell'unità di percorso è allora[44] $\sim g_0 \approx 68,5 e \sim 2 GeV/kgauss \cdot m$. Risulta pertanto che il campo magnetico necessario per accelerare un monopolo è almeno $\sim \rho_{(aria)} \cdot 4700 \mathcal{H}/g_0 \sim$ ½ *kgauss*, ovvero ~ 1000 volte il valore del campo magnetico terrestre ( ~ ½ gauss). Se ne conclude che i monopoli dovrebbero rimanere intrappolati nelle linee di forza. In questa situazione i monopoli (e gli antimonopoli) andrebbero ad accumularsi ai poli terrestri della polarità opposta, finendo per neutralizzare l'effetto del magnetismo terrestre. Dal momento che ciò non è finora accaduto, o i monopoli non esistono, oppure non arrivano, se non in numero assolutamente irrilevante, fino all'atmosfera terrestre.

### *La storia continua*

Nell'ultima parte dell'articolo del 1931 Dirac sviluppò i particolari matematici della teoria dell'elettrone nel campo di un polo isolato (e, reciprocamente, di un monopolo nel campo di una carica), che generalizzò successivamente (1948) a un sistema dinamico di cariche e poli. Notò con un certo disappunto che la teoria non forniva, come egli sperava, una relazione diretta tra il valore della carica dell'elettrone e le costanti fondamentali $h$ e $c$, ma portava alla condizione di quantizzazione (1), che stabilisce soltanto un legame di reciprocità tra i quanti di carica elettrica e magnetica. Concluse però, con enfasi: «La reciprocità tra elettricità e magnetismo è perfetta».

La straordinaria intuizione di Dirac del monopolo ha generato molti sviluppi interessanti[45]. Quando l'elettromagnetismo viene generalizzato, per descrivere le interazioni nucleari deboli e forti, con il formalismo delle teorie di *gauge* o teorie di Yang-Mills, si trovano stati eccitati dei campi che appaiono a grandi distanze come monopoli magnetici. In questa loro incarnazione moderna i monopoli non hanno una linea nodale, o stringa di Dirac, che li accompagna, e ciò ne facilita in qualche modo l'accettazione da parte dei fisici. I monopoli fanno la loro comparsa anche nelle cosiddette *Teorie di Grande Unificazione* (*GUT*), che descrivono l'emergere delle forze fondamentali nell'universo primordiale. Queste teorie prevedono che i monopoli si siano formati nel momento in cui l'elettromagnetismo si separò dalle altre forze. Tuttavia le previsioni indicano che questi monopoli erano in numero molto limitato e avevano una distribuzione uniforme nell'universo: proprio quella richiesta per spiegare perché i monopoli non vengono osservati! Negli ultimi anni le previsioni teoriche della loro comparsa si sono moltiplicate, al punto che i monopoli vengono oggi considerati un naturale prodotto di qualsiasi teoria di campo, e non solo dell'elettromagnetismo. Essi compaiono anche nella teoria delle stringhe, come *brane* con carica magnetica, e nella superconduttività, come flussi magnetici quantizzati. In definitiva, i monopoli sembrano giocare un ruolo fondamentale nell'odierna concezione della natura.

Con tutto ciò resta il fatto che monopoli magnetici non ne sono mai stati trovati, né di antichi né di moderni, e forse non se ne troveranno mai. Che cosa pensava Dirac al riguardo? Nell'ultimo periodo della sua vita (morì nel 1984 - era nato nel 1902), ormai deluso dalla mancanza di conferme sperimentali in tutti quegli anni, espresse l'opinione che i monopoli magnetici non esistono (era il 1981).

Chi ha ragione, il Dirac del 1931 o quello del 1981? Non siamo ancora in grado di rispondere a questa domanda. In ogni caso, la storia continua.

---

[1] La carica (o massa) magnetica di un polo magnetico è definita come $g=F/H$, dove $F$ è la forza esercitata sul polo da un campo magnetico esterno $H$.

[2] Il potenziale (elettrostatico) di una carica isolata $q$ è dato dallo scalare $V = -q/r$, definito ovunque a meno di una costante ininfluente. Tale potenziale è singolare ($V = \infty$) nell'origine ($r=0$). Il potenziale scalare è connesso al carattere irrotazionale del campo elettrostatico: $rot\mathbf{E} = 0$ implica infatti che il campo elettrico $\mathbf{E}$ è il gradiente di uno scalare ($\mathbf{E} = gradV$).

[3] Il potenziale vettore magnetico di un monopolo di carica magnetica $g$, espresso in coordinate polari sferiche, è un vettore di modulo $A = A_\varphi = \dfrac{g}{r}\left(\dfrac{1-\cos\theta}{\sin\theta}\right)$ diretto lungo $\hat{\varphi}$, infinito nei punti dove $\theta=\pi$ (semiasse $z$-negativo). Il potenziale magnetico si ricava formalmente come soluzione dell'equazione $\mathbf{H} = rot\mathbf{A}$, che in questo caso si riduce a $H = H_r = \dfrac{g}{r^2} = \dfrac{1}{r\sin\theta}\cdot\left[\dfrac{\partial}{\partial\theta}(A_\varphi\sin\theta)-\dfrac{\partial}{\partial\varphi}A_\theta\right]$, essendo $H_\theta = \dfrac{1}{r\sin\theta}\cdot\left[\dfrac{\partial A_r}{\partial\varphi}-\dfrac{\partial}{\partial r}\left(r\sin\theta A_\varphi\right)\right] = 0$ e $H_\varphi = \dfrac{1}{r}\cdot\left[\dfrac{\partial}{\partial r}(rA_\theta)-\dfrac{\partial A_r}{\partial\theta}\right] = 0$. La nostra dimostrazione non fa uso di queste equazioni. Il potenziale vettore magnetico $\mathbf{A}$ è definito a meno del gradiente di una funzione scalare e soddisfa l'equazione $div\mathbf{A} = 0$. Il campo magnetico ha una singolarità ($H = \infty$) nell'origine ($r=0$).

[4] Come dimostra l'effetto Aharonov-Bohm, il potenziale vettore magnetico $\mathbf{A}$ può modificare la fase della funzione d'onda anche in regioni, come l'esterno di un solenoide, dove il campo magnetico è nullo. Per questo fatto $\mathbf{A}$, e non $(\mathbf{E},\mathbf{H})$, può essere considerato il "vero" campo. Sul significato fisico del potenziale vettore magnetico in meccanica quantistica v. la discussione in: R.P. Feynman, in «La Fisica di Feynman», II (Elettromagnetismo e materia), Zanichelli (2001).

[5] Una di tali proprietà "esotiche" dei monopoli è la violazione della parità nell'interazione tra monopoli e cariche elettriche [v. voce *Monopolo magnetico*, di A.S. Goldhaber, in «Enciclopedia della Scienza e della Tecnica Mondadori-McGraw-Hill», VIII, p. 639 (1980)].

[6] Come si verifica facilmente: $\int_{-\infty}^{+\infty}\left(\lim_{a\to 0}d_a(\rho)\right)d\rho = \lim_{a\to 0}\int_{-a}^{+a}\dfrac{1}{2a}d\rho = \lim_{a\to 0}\left(\dfrac{1}{2a}\cdot 2a\right) = 1$, da confrontare con $\int_{-\infty}^{+\infty}\delta(\rho)d\rho = 1$.

[7] Il campo all'esterno della stringa ($[|x|+|y| > 0, z<0] \cup [|x|+|y| \geq 0, z>0]$) si annulla: $H_z(\text{ext}) = 8g(1-\dfrac{z}{|z|})\cdot\left.\left[\delta(\sqrt{x^2+y^2})\right]^2\right|_{|x|+|y|\geq 0} = 0$. A rigore si dovrebbe esprimere a grandi distanze ($|z| \gg a$) la discontinuità tra il campo radiale all'interno del semi-solenoide e il campo longitudinale all'esterno mediante la funzione "scalino" $\theta(x)$ di Heaviside ($\theta(x)=0$ per $x<0$, $\theta(x)=1$ per $x>0$): $\mathbf{H} \cong H_r\left[1-\theta\left(a-\sqrt{x^2+y^2}\right)\cdot\theta(-z)\right]\hat{r} + H_z\cdot\theta\left(a-\sqrt{x^2+y^2}\right)\cdot\theta(-z)\hat{z}$. Per $a\to 0$ si ottiene il campo a tutte le distanze: $\lim_{a\to 0}\mathbf{H} = \dfrac{g}{r^2}\left[1-\theta\left(-\sqrt{x^2+y^2}\right)\cdot\theta(-z)\right]\hat{r} + 8g\left(1-\dfrac{z}{|z|}\right)\theta\left(-\sqrt{x^2+y^2}\right)\cdot\theta(-z)\left[\delta(\sqrt{x^2+y^2})\right]^2\hat{z}$ (si noti che sul semiasse $z$-positivo il campo si annulla come $0^3\cdot\infty^2$).

[8] Si noti che $H_z < H_M$ per $z>0$ e viceversa, con il valore massimo per $H_z(z = -a/\sqrt{15}) = \dfrac{5}{4}H_M$ [v. *Elettricità e magnetismo*, di E.M. Purcell, in «La Fisica di Berkeley», II, parte II, Zanichelli (1971)].

[9] Sono state escogitate varie tecniche per aggirare la singolarità del potenziale vettore magnetico. Una di queste sfrutta la proprietà d'invarianza di *gauge* del campo elettromagnetico: il potenziale viene definito come $\mathbf{A}_1 = \dfrac{g}{r}\left(\dfrac{1-\cos\theta}{\sin\theta}\right)\hat{\varphi}$ ovunque tranne in un cono di piccola apertura con vertice in $M$ (*Fig. 1*), e come $\mathbf{A}_2 = -\dfrac{g}{r}\left(\dfrac{1+\cos\theta}{\sin\theta}\right)\hat{\varphi}$ all'interno di questo cono. Nessuno di questi due potenziali è singolare nella regione in cui è definito. Ciascuno si ottiene dall'altro con una trasformazione di *gauge* che lascia invariato il campo elettromagnetico, p. es. $\mathbf{A}_2 = \mathbf{A}_1 - \dfrac{2g}{r\sin\theta}\hat{\varphi}$. Il termine di *gauge* $-\dfrac{2g}{r\sin\theta}\hat{\varphi}$ è il gradiente di una funzione scalare, in questo caso $-2g\varphi$ $\left[\text{prova}: \nabla(-2g\varphi) = \dfrac{1}{r\sin\theta}\dfrac{\partial(-2g\varphi)}{\partial\varphi}\hat{\varphi} = -\dfrac{2g}{r\sin\theta}\hat{\varphi}\right]$.

[10] P.A.M. Dirac, *Quantized singularities in the electromagnetic field*, in «Proceedings of the Royal Society», A133, p. 60 (1931). In seguito pubblicò un altro importante articolo sulla teoria dei poli magnetici: P.A.M. Dirac, *The Theory of Magnetic Poles*, in «Phys. Rev.», 74, 7 (1948).

[11] Questa forma della condizione di quantizzazione, dove figura una carica $q$ qualsiasi, è più generale dell'originaria forma con la carica $e$ dell'elettrone, valida per l'interazione elettrone-monopolo considerata da Dirac. A causa di questa "elasticità" del valore della carica la ricerca sperimentale di monopoli rischia di non avere mai termine [v. A. Loinger,

*Sul monopolo magnetico di Dirac*, in «Il Nuovo Saggiatore», 19, n. 5-6, p. 3 (2003); v. anche su http://arXiv.org/physics/0309091].

[12] Ricordiamo che l'equazione d'onda $i\hbar\frac{\partial \psi}{\partial t} = \hat{H}\psi$ ammette soluzioni della forma $\psi = C \cdot \psi_0$, dove $\psi_0$ è una funzione d'onda ordinaria (cioè a un sol valore in ogni punto) e $C$ un coefficiente costante arbitrario, in generale complesso della forma $|C| \cdot e^{i\alpha}$. Imponendo, se l'integrale converge, la condizione di normalizzazione della funzione d'onda $\int |\psi|^2 \cdot dV = 1$, $|C|$ viene reso univoco facendogli assumere un valore unitario e si può esprimere $\psi$, che resta non univoca per l'arbitrarietà della scelta di $\alpha$, nella forma $\psi_0 \cdot e^{i\alpha}$. Sotto questa forma la non univocità della funzione d'onda viene interamente confinata nel fattore di fase indeterminato $e^{i\alpha}$. Quando l'integrale $\int |\psi|^2 \cdot dV$ diverge, $|\psi|^2$ non dà i valori assoluti della probabilità delle coordinate, ma si può dire soltanto che è proporzionale a questa probabilità. In tali casi si considera il rapporto tra $|\psi|^2$ in due diversi punti, che determina la probabilità relativa delle coordinate di questi punti ed è indipendente da $C$ [v. L.D. Landau - E.M. Lifšits, *Meccanica Quantistica*, Editori Riuniti (1976)]. Si dimostra senza difficoltà che questa indeterminazione di fase non influisce su nessun risultato fisico. Infatti i fattori di fase indeterminati si cancellano reciprocamente nei prodotti $\phi\psi = \psi^*\psi = |\psi|^2$ che intervengono nelle operazioni più semplici, mentre nelle operazioni più generali con prodotti tipo $\phi_m\psi_n$ residuano fattori di fase della forma $e^{i(\alpha_n - \alpha_m)}$, che scompaiono calcolando il quadrato del modulo dell'integrale $\int \psi_m^* \psi_n dV$. Così avviene anche per altri tipi di operazioni generali, quali trasformazioni fra rappresentazioni, elementi di matrice, ecc. (v. note [18] e [19]).

[13] Naturalmente non stiamo parlando di traiettorie reali percorse da un corpo in un determinato tempo, ma di curve d'integrazione (in generale in quattro dimensioni) lungo le quali vengono calcolati degli integrali di linea. Con riferimento alla *Fig. 3* abbiamo: (lungo $C_1$) $\Delta\varphi_1 = \int_A^B \nabla\varphi \cdot ds_1$, (lungo $C_2$) $\Delta\varphi_2 = \int_A^B \nabla\varphi \cdot ds_2$. Intorno alla curva chiusa ($C = C_1 - C_2$) abbiamo quindi:

$$\beta = \oint \nabla\varphi \cdot ds = \int_A^B \nabla\varphi \cdot ds_1 + \int_B^A \nabla\varphi \cdot ds_2 = \int_A^B \nabla\varphi \cdot ds_1 - \int_A^B \nabla\varphi \cdot ds_2 = \Delta\varphi_1 - \Delta\varphi_2.$$

[14] $\oint_C \nabla\gamma \cdot ds = 0$ per ipotesi e $\oint_C \nabla\alpha \cdot ds \equiv 0$ per l'identità $\nabla\alpha \equiv 0$.

[15] Affinché sia lecito esprimere $\psi$ in questa forma è necessario verificare che $\psi_1$ sia anch'essa una funzione d'onda, ossia soddisfi un'equazione d'onda della forma $\hat{H}_1\psi_1 = \hat{W}_1\psi_1$ quando $\psi$ soddisfa contestualmente l'equazione $\hat{H}\psi = \hat{W}\psi$ (supponiamo che le equazioni siano nella forma non relativistica). Tale verifica è piuttosto semplice nel caso di una particella libera (p.es. elettrone non soggetto a forze esterne). In tal caso l'operatore hamiltoniano $\hat{H}$ si riduce al termine cinetico $\frac{1}{2m}\hat{\mathbf{p}}^2$, proporzionale al quadrato dell'operatore quantità di moto $\hat{\mathbf{p}} = -i\hbar\nabla$, mentre $\hat{W}$ rappresenta l'operatore energia $i\hbar\frac{\partial}{\partial t}$. Applicando questi operatori alla $\psi$ si trova

$\hat{\mathbf{p}}\psi = -i\hbar\nabla(e^{i\beta}\psi_1) = e^{i\beta}(-i\hbar\nabla + \hbar\nabla\beta)\psi_1 = e^{i\beta}(\hat{\mathbf{p}} + \hbar\boldsymbol{\kappa})\psi_1$, dove $\boldsymbol{\kappa} = \nabla\beta$, e

$i\hbar\frac{\partial}{\partial t}\psi = e^{i\beta}(i\hbar\frac{\partial}{\partial t} - \hbar\frac{\partial\beta}{\partial t})\psi_1 = e^{i\beta}(\hat{W} - \hbar\kappa_0)\psi_1$, dove $\kappa_0 = \frac{\partial\beta}{\partial t}$.

Introducendo i nuovi operatori $\hat{\mathbf{p}}_1 = \hat{\mathbf{p}} + \hbar\boldsymbol{\kappa}$ e $\hat{W}_1 = \hat{W} - \hbar\kappa_0$, e applicando nuovamente alla $\psi$ l'operatore quantità di moto, si ottiene $\hat{\mathbf{p}}^2\psi = e^{i\beta}\hat{\mathbf{p}}_1^2\psi_1$, da cui è facile verificare che $\psi_1$ soddisfa l'equazione $\hat{H}_1\psi_1 = \hat{W}_1\psi_1$ con $\hat{H}_1 = \frac{1}{2m}\hat{\mathbf{p}}_1^2$.

Nel caso relativistico si arriva senza difficoltà allo stesso risultato esprimendo l'equazione d'onda come nella forma non relativistica (valida se la particella ha spin diverso da zero). Per verificarlo partiamo dall'equazione relativistica (e. di Dirac) $(\alpha p - m_0 c\beta)\psi = 0$, dove $\alpha = (\beta, \hat{\alpha})$ e $\hat{\alpha}$ e $\beta$ sono le matrici

$$\alpha^1 = \begin{vmatrix} 0 & 0 & 0 & 1 \\ 0 & 0 & 1 & 0 \\ 0 & 1 & 0 & 0 \\ 1 & 0 & 0 & 0 \end{vmatrix}, \quad \alpha^2 = \begin{vmatrix} 0 & 0 & 0 & -i \\ 0 & 0 & i & 0 \\ 0 & -i & 0 & 0 \\ i & 0 & 0 & 0 \end{vmatrix}, \quad \alpha^3 = \begin{vmatrix} 0 & 0 & 1 & 0 \\ 0 & 0 & 0 & -1 \\ 1 & 0 & 0 & 0 \\ 0 & -1 & 0 & 0 \end{vmatrix}, \quad \beta = \begin{vmatrix} 1 & 0 & 0 & 0 \\ 0 & 1 & 0 & 0 \\ 0 & 0 & -1 & 0 \\ 0 & 0 & 0 & -1 \end{vmatrix}$$, $p$ l'operatore $p = (i\frac{\hbar}{c}\frac{\partial}{\partial t}, \hat{\mathbf{p}})$, $m_0$ la massa

di riposo della particella e $c$ la velocità della luce. Allora $\alpha p = i\frac{\hbar}{c}\beta\frac{\partial}{\partial t} - \alpha\hat{\mathbf{p}}$. Moltiplicando a sinistra per $\beta$, e tenendo conto che $\beta^2 = 1$, si può riscrivere l'equazione d'onda relativistica nella forma $i\hbar\frac{\partial}{\partial t} = H\psi$ con $H = \beta c(\alpha\hat{\mathbf{p}} + m_0 c)$ e ricavare, come prima, $H_1 \psi_1 = W_1 \psi_1$ con $H_1 = \beta c(\alpha\hat{\mathbf{p}}_1 + m_0 c)$ e $W_1 = i\hbar\frac{\partial}{\partial t}$. Si vedrà più avanti l'interpretazione fisica di questi risultati.

[16] La funzione densità rappresenta la densità di probabilità delle coordinate.

[17] Il quadrato del modulo dell'integrale $\int \phi_m \psi_n dx dy dz$, che è connesso all'ampiezza di sovrapposizione delle due funzioni d'onda $\psi_m, \psi_n$, viene interpretato fisicamente come la probabilità di accordo dei due stati $m$ ed $n$, ovvero come la probabilità di trovare $m$ in $n$.

[18] In una rappresentazione diagonale nelle coordinate, consideriamo i vettori di base (autovettori delle coordinate) corrispondenti ai punti $A$ e $B$ di *Fig. 3*, presi nello stesso istante $t$. I vettori $|A\rangle$ e $|A'\rangle = e^{i\beta'}|A\rangle$ descrivono lo stesso stato fisico, e così pure i vettori $|B\rangle$ e $|B''\rangle = e^{i\beta''}|B\rangle$. In questa rappresentazione $|A'\rangle \equiv |x'y'z't\rangle = |\mathbf{r}'\rangle = \delta(\mathbf{r}-\mathbf{r}') \cdot e^{i\beta'}$ e $|B''\rangle \equiv |x''y''z''t\rangle = |\mathbf{r}''\rangle = \delta(\mathbf{r}-\mathbf{r}'') \cdot e^{i\beta''}$. Otteniamo allora per gli elementi di matrice delle coordinate $\langle \mathbf{r}'|\hat{\mathbf{r}}|\mathbf{r}''\rangle = \int \delta(\mathbf{r}-\mathbf{r}') \cdot e^{-i\beta'} \mathbf{r} \delta(\mathbf{r}-\mathbf{r}'') \cdot e^{i\beta''} d\mathbf{r} = \mathbf{r}'\delta(\mathbf{r}'-\mathbf{r}'')$. Nella stessa rappresentazione il vettore corrispondente a una generica funzione d'onda $\psi_n$ è $|\psi_n\rangle = \int \psi_{1n}(\mathbf{r}')|\mathbf{r}'\rangle d\mathbf{r}' = \int \psi_{1n}(\mathbf{r}')\delta(\mathbf{r}-\mathbf{r}') \cdot e^{i\beta'} d\mathbf{r}' = \psi_{1n}(\mathbf{r}) \cdot e^{i\beta}$, che corrisponde alla funzione d'onda stessa nella rappresentazione delle coordinate (rappresentata dal vettore che ha per componenti i valori della funzione per tutti i valori delle coordinate). Inoltre la componente $r$-esima del vettore di stato è $\langle \mathbf{r}|\psi_n\rangle = \psi_{1n}(\mathbf{r})$, cioè il valore particolare della funzione d'onda nel punto di coordinate $\mathbf{r}$. Ciò premesso, in una seconda rappresentazione, in cui le osservabili $\xi$ sono diagonali, abbiamo $|\psi_n\rangle = \sum_{\xi'} \varphi(\xi')|\xi'\rangle$, dove $\varphi(\xi) = \langle \xi|\psi_n\rangle = \langle \xi|\int \psi_{1n}(\mathbf{r}')|\mathbf{r}'\rangle d\mathbf{r}' = \int \langle \xi|\mathbf{r}'\rangle \psi_{1n}(\mathbf{r}') d\mathbf{r}'$. Le funzioni di trasformazione (trasformate di Fourier generalizzate) che consentono di passare dalla prima alla seconda rappresentazione sono date da $\langle \xi|\mathbf{r}'\rangle = \int \xi^*(\mathbf{r})\delta(\mathbf{r}-\mathbf{r}') \cdot e^{i\beta'} d\mathbf{r} = \xi^*(\mathbf{r}') \cdot e^{i\beta'}$, da cui si ricava $\varphi(\xi) = \int \xi^*(\mathbf{r}') \cdot \psi_1(\mathbf{r}') e^{i\beta'} d\mathbf{r}'$. La funzione d'onda trasformata $\varphi$ ha la fase determinata (a meno di una costante ininfluente indipendente da $\xi$), perciò $\xi^*$ ha la forma di una $\phi$, e quindi $\xi$ ha un'indeterminazione di fase $-\beta$ (a meno di tale costante).

[19] Esprimendo il prodotto della matrice $\hat{\alpha}$ per la funzione d'onda $\psi_n$ nella forma $|\hat{\alpha}\psi_n\rangle = \iint |\mathbf{r}'\rangle \langle \mathbf{r}'|\hat{\alpha}|\mathbf{r}''\rangle \langle \mathbf{r}''|\psi_n\rangle d\mathbf{r}' d\mathbf{r}''$ abbiamo, nella rappresentazione delle coordinate, $\langle \mathbf{r}''|\psi_n\rangle = \int \delta(\mathbf{r}-\mathbf{r}'') \cdot e^{-i\beta(\mathbf{r}'')-i\alpha} \cdot \psi_n(\mathbf{r}) d\mathbf{r} = e^{-i\beta''} \psi_n(\mathbf{r}'') = e^{-i\beta''} \psi_{1n}(\mathbf{r}'') \cdot e^{i\beta''} = \psi_{1n}(\mathbf{r}'')$, $\langle \mathbf{r}'|\hat{\alpha}|\mathbf{r}''\rangle = \alpha\delta(\mathbf{r}'-\mathbf{r}'')$ e $|\mathbf{r}'\rangle = \delta(\mathbf{r}-\mathbf{r}') \cdot e^{i\beta'}$, da cui $|\hat{\alpha}\psi_n\rangle = \iint \delta(\mathbf{r}-\mathbf{r}') \cdot e^{i\beta'} \hat{\alpha}\delta(\mathbf{r}'-\mathbf{r}'') \psi_{1n}(\mathbf{r}'') d\mathbf{r}' d\mathbf{r}'' = e^{i\beta} \cdot \hat{\alpha}\psi_1(\mathbf{r})$.

[20] Se per due 4-vettori $k, k'$ $\oiint_S rot k' \, dS = \oiint_S rot k \, dS$ per una stessa superficie arbitraria $S$ delimitata dalla curva $C$, allora $\oiint_S (rot k' - rot k) dS = \oiint_S rot(k'-k) dS = 0$, che implica $\oint_C (k'-k) \cdot ds = 0$. Questo significa che $k'-k$ è integrabile, dunque uguale per definizione al gradiente di una funzione scalare (a un sol valore) $\chi$: $k'-k = grad\chi$. Ma $rot(grad\chi) = 0$, da cui $rot k' = rot k$.

[21] Infatti in questo caso $\beta_{el} = \oint k_0 dt = k_0 \oint dt \equiv 0$.

[22] H. Weyl, «Z. Physik», 56, 330 (1929). L'invarianza di *gauge* viene chiamata anche invarianza elettromagnetica o (con termine tedesco) *Eichinvarianz*. Vediamo brevemente in che cosa consiste. In quattro dimensioni è possibile esprimere matematicamente le equazioni dell'elettromagnetismo in una notazione molto stringata:

$\Box A_\mu = \nabla_\mu \nabla_\mu A_\mu = j_\mu$, $\nabla_\mu j_\mu = 0$, dove $A_\mu$ è la componente $\mu$–esima del *4*-vettore potenziale elettromagnetico $A = (A_0, \mathbf{A})$, $j_\mu$ il *4*-vettore densità di corrente $j_\mu = (\rho, \frac{1}{c}\mathbf{j})$, $\nabla_\mu$ la componente $\mu$–esima dell'operatore gradiente $\nabla = (\frac{1}{c}\frac{\partial}{\partial t}, -\tilde{\nabla}) = (\frac{1}{c}\frac{\partial}{\partial t}, -\frac{\partial}{\partial x}, -\frac{\partial}{\partial y}, -\frac{\partial}{\partial z})$, $\tilde{\nabla}^2$ il laplaciano, $\Box = \nabla_\mu \nabla_\mu$ l'operatore dalambertiano

$\Box = \nabla_\mu \nabla_\mu = \frac{1}{c^2}\frac{\partial^2}{\partial t^2} - \tilde{\nabla}^2 = \frac{1}{c^2}\frac{\partial^2}{\partial t^2} - \frac{\partial^2}{\partial x^2} - \frac{\partial^2}{\partial y^2} - \frac{\partial^2}{\partial z^2}$. Nella notazione vettoriale tridimensionale le equazioni precedenti diventano: $\Box A_0 = -\rho$, $\Box \mathbf{A} = -\frac{1}{c}\mathbf{j}$.

In aggiunta a queste equazioni viene richiesta la condizione di Lorentz $\nabla_\mu A_\mu = 0$ (in notazione tridimensionale $\frac{1}{c}\frac{\partial A_0}{\partial t} + \tilde{\nabla}\cdot \mathbf{A} = 0$), per garantire la forma relativisticamente invariante delle equazioni di Maxwell.

Dal potenziale elettromagnetico si ricavano i campi elettrico $\mathbf{E} = -\frac{1}{c}\frac{\partial \mathbf{A}}{\partial t} - \tilde{\nabla}A_0$ e magnetico $\mathbf{H} = \tilde{\nabla}\wedge \mathbf{A}$ (o $\mathbf{H} = rot\mathbf{A}$).

La trasformazione di gauge del campo elettromagnetico è $A'_\mu = A_\mu + \nabla_\mu \chi$ ovvero $A'_0 = A_0 + \frac{1}{c}\frac{\partial \chi}{\partial t}$, $\mathbf{A}' = \mathbf{A} - \tilde{\nabla}\chi$, dove $\chi$ è una funzione arbitraria che soddisfa la condizione $\Box \chi = 0$ (equivalente alla condizione di Lorentz).
La verifica dell'invarianza elettromagnetica per trasformazioni di *gauge* è immediata:
$\nabla_\mu A'_\mu = \nabla_\mu A + \nabla_\mu \nabla_\mu \chi = \nabla_\mu A + \Box \chi = \nabla_\mu A_\mu$, da cui segue l'invarianza delle equazioni del campo elettromagnetico.

L'invarianza di *gauge* richiede la contestuale trasformazione della funzione d'onda $\psi' = \psi e^{i\frac{q}{\hbar c}\chi}$ e degli operatori $\hat{p}' = \hat{p}_1 + \frac{q}{c}\tilde{\nabla}\chi$, $\hat{W}' = \hat{W}_1 - q\frac{\partial \chi}{\partial t}$, com'è facile verificare (v. nota[15]). Le possibili funzioni d'onda di una particella in un campo elettromagnetico differiscono per una trasformazione di *gauge*.

[23] $e^{i\varphi + 2\pi ni} \equiv e^{i\varphi}$ per qualsiasi intero $n$, come si ricava dalla formula di Eulero $e^{i\pi} = -1 \rightarrow e^{2\pi ni} = 1$.

[24] L'equazione d'onda è un'equazione differenziale del second'ordine in cui figurano le derivate seconde della funzione d'onda, che devono esser definite. Affinché le derivate seconde siano definite devono esser continue sia le derivate prime che la funzione d'onda medesima.

[25] Quando la funzione d'onda s'annulla ($\psi = |\psi|e^{i\varphi} = 0$) l'ampiezza è nulla ($|\psi| = 0$) mentre il fattore di fase ($e^{i\varphi}$) può assumere un valore qualsiasi, di modo che la fase è totalmente indeterminata. La funzione d'onda resta comunque continua perché il valore nullo dell'ampiezza cancella la discontinuità della fase, dando un prodotto continuo. Sebbene la fase possa assumere su una linea nodale qualsiasi valore, la teoria di Dirac richiede che possa acquisire soltanto incrementi banali, cioè multipli interi di $2\pi$.

[26] Scrivendo la funzione d'onda come $\psi(xyzt) = u(xyzt) + i\cdot v(xyzt)$ il suo annullarsi ($\psi = 0$) implica le due condizioni
$\begin{cases} u(xyzt) = 0 \\ v(xyzt) = 0 \end{cases}$
che rappresentano le equazioni di due (iper)superfici tridimensionali immerse nello spazio quadridimensionale. Le due (iper)superfici si intersecano in generale su una (iper)superficie nodale bidimensionale, allo stesso modo in cui due superfici bidimensionali immerse nello spazio tridimensionale si intersecano su una linea nodale. Le (iper)superfici nodali bidimensionali possono essere circondate da curve (in quattro dimensioni) nello stesso modo in cui le linee nodali possono essere circondate da curve in tre dimensioni.

[27] Consideriamo infatti il cambiamento di fase di due funzioni d'onda intorno ad una stessa curva chiusa molto piccola. La prima funzione d'onda, che non ha una linea nodale passante attraverso la curva considerata, ha un cambiamento di fase $\beta$ molto piccolo ($\beta \ll 2\pi$). La seconda funzione d'onda, che ha una linea nodale che attraversa la stessa curva, ha allora un cambiamento di fase $\beta + 2\pi n$, che è all'incirca multiplo di $2\pi$ ($\beta + 2\pi n \simeq 2\pi n$).

[28] v. nota precedente.

[29] Una linea nodale definisce un asse privilegiato (supponiamo sia il semiasse $z$-negativo). Intorno a quest'asse di simmetria: $k_r = k_\theta = A_r = A_\theta = 0$ e $k = k_\varphi, A = A_\varphi$, che dipendono da $r$ e $\theta$ (ma non da $\varphi$). Con questa scelta: $H_\theta = H_\varphi = 0$ e $H = H_r$, che dipende solo da $r$ (v. nota [3]). Allora il cambiamento di fase intorno a una curva chiusa molto piccola che circonda la linea nodale diventa:

$$\beta = \oint k_\varphi r \sin\theta d\varphi \simeq 2\pi k_\varphi r \sin\theta = 2\pi \frac{q}{\hbar c} A_\varphi r \sin\theta = 2\pi n + \iint_S \frac{q}{\hbar c} \mathbf{H} \cdot d\mathbf{S} \simeq 2\pi n - \frac{q}{\hbar c} H_r 2\pi r^2 (1+\cos\theta)$$, da cui

$A_\varphi r \sin\theta + H_r r^2 (1+\cos\theta) = n\frac{\hbar c}{q}$, che trasformiamo in $\left[ A_\varphi r \sin\theta - H_r r^2 (1-\cos\theta) \right] + 2H_r r^2 = n\frac{\hbar c}{q}$. Poiché solo il termine entro parentesi quadre dipende da $\theta$, ed il secondo membro non dipende né da $\theta$ né da $r$, allora $H_r r^2$ =costante=$g$ e $A_\varphi r \sin\theta - H_r r^2(1-\cos\theta) = A_\varphi r \sin\theta - g(1-\cos\theta) = 0$, da cui infine ritroviamo $A_\varphi = g\frac{(1-\cos\theta)}{r\sin\theta}$, che ha la singolarità richiesta per $\theta = \pi$.

[30] Infatti $\Phi_H = H_r \cdot 4\pi r^2$ e, per la nota precedente, $H_r = \frac{g}{r^2}$, da cui $\Phi_H = 4\pi g$.

[31] E. Fermi, *Conferenze di Fisica Atomica*, in *Note e memorie (Collected papers)*, vol. II (Accademia Nazionale dei Lincei, The University of Chicago Press), 1966, p. 684. Dopo questa conferenza apparve suun noto quotidiano un articolo dal titolo "Conferenza di Fermi sui Monopoli", che suscitò la bonaria ironia di Fermi per l'evidente fraintendimento (…che Fermi avesse parlato dei Monopoli di Stato!?).

[32] P.B. Price et al., «Phys. Rev. Lett.», 35, 486 (1975).

[33] N. Bohr, «Phil. Mag.», 25, 10 (1913); ibid., 30, 581 (1915).

[34] Cioè la teoria è inapplicabile per $\Delta v_e \sim v$.

[35] L'impulso longitudinale è nullo: $\Delta p_\parallel = eE_{max}^{(Z_1)} \cdot \left(\frac{b}{v}\right) \cdot \int_{-\infty}^{+\infty} \frac{\xi}{\left(1+\xi^2\right)^{\frac{3}{2}}} d\xi = eE_{max}^{(Z_1)} \cdot \left(\frac{b}{v}\right) \cdot \left[-\frac{1}{\sqrt{1+\xi^2}}\right]_{-\infty}^{+\infty} = 0$.

[36] Per l'aria in condizioni normali $N \approx 2,5 \cdot 10^{19} cm^{-3}$, $\mathscr{N} \approx 3,8 \cdot 10^{20} cm^{-3}$.

[37] Poiché i parametri di collisione compaiono in un termine logaritmico è sufficiente stimarne l'ordine di grandezza. Il semidiametro di collisione $b_{min}$, cioè la distanza di massimo avvicinamento, corrisponde al massimo trasferimento d'energia in una collisione frontale elastica nella diffusione Rutherford tra due particelle cariche, dato classicamente da $\Delta W_{max} = \frac{2mM^2}{(m+M)^2} v^2$, che (essendo $M \gg m$) si può approssimare con $\sim 2mv^2$. Nella collisione l'elettrone rincula con quantità di moto $\sim 2mv$. Lo stesso risultato si può ottenere semplicemente dalla conservazione dell'energia e della quantità di moto in una collisione elastica con deflessione del proiettile a $180°$. Confrontando $\Delta W_{max}$ con $\Delta W = \frac{2Q^2 e^2}{mb^2 v^2}$ si ottiene $b_{min} \sim \frac{Qe}{mv^2}$. Il raggio adiabatico $b_{max}$ si stima considerando che a grande distanza le collisioni sono poco intense e di lunga durata. In queste condizioni si può assumere che l'impulso conferito all'elettrone sia uguale al prodotto della forza impulsiva $eE_{max}$ per la durata dell'impulso $t = 2b/v$. Quando la durata della collisione $t$ è molto maggiore del periodo orbitale $T$ dell'elettrone ($t \gg T$) la forza impulsiva è molto più piccola della forza che mantiene l'elettrone nella sua orbita e l'orbita viene perturbata adiabaticamente, sicché non v'è praticamente trasferimento d'energia all'elettrone. Stimando in $t \sim b/v$ la durata della collisione, questo significa che l'elettrone assorbe energia soltanto quando $t \sim T$, cioè $b_{max} \sim v \cdot T = \frac{v}{\omega}$, dove $\omega$ è la frequenza orbitale dell'elettrone.

[38] La versione relativistica della formula di Bohr per una particella carica è $\mathscr{F}_{Z_1} = \frac{4\pi mc^2 r_e^2 Z_1^2 \mathscr{N}}{\beta^2} \ln\left(\frac{\gamma^2 mv^3}{\omega Z_1 e^2}\right)$, dove $\gamma^2 = \frac{1}{1-\beta^2}$, e $\mathscr{F}_{g_0} \approx 4\pi mc^2 r_e^2 \mathscr{N} \cdot 4700 \cdot \ln\left(\frac{2\gamma^2 mv^2 c}{\omega \cdot 137 e^2}\right)$ per il monopolo. Le versioni quantistiche sono rispettivamente $\mathscr{F}_{Z_1} = \frac{4\pi mc^2 r_e^2 Z_1^2 \mathscr{N}}{\beta^2} \ln\left(\frac{\gamma^2 mv^2/2}{\hbar v}\right)$ e $\mathscr{F}_{g_0} \approx 4\pi mc^2 r_e^2 \mathscr{N} \cdot 4700 \cdot \ln\left(\frac{\gamma^2 mv^2/2}{\hbar v}\right)$. Queste formule si ricavano ricordando che dal punto di vista quantistico hanno significato solo le lunghezze superiori alla lunghezza d'onda dell'elettrone $\lambdabar = \hbar/p$, da cui $b_{min} \sim \lambdabar = \hbar/p \sim \frac{\hbar}{\gamma mv}$; inoltre, da $t = 2b/v$ segue $b_{max} = \frac{v}{2v/\gamma}$. Una più precisa trattazione relativistica dovuta a Bethe fornisce per le particelle $\mathscr{F}_{Z_1} = \frac{4\pi mc^2 r_e^2 Z_1^2 \mathscr{N}}{\beta^2} \left\{ \ln\left(\frac{2mc^2 \gamma^2 \beta^2}{\bar{I}}\right) - \beta^2 \right\}$, dove $\bar{I}$ è il

potenziale di ionizzazione medio degli atomi del mezzo [per l'aria $\overline{I} \simeq 12 eV$, e in generale $\overline{I} \simeq 13,6 Z_2 \, eV$ per $Z_2 > 16$ (Zolfo)]. Si noti che in queste formule la perdita d'energia non dipende dalla massa della particella. Si può esprimere il potere frenante anche in funzione dell'energia della particella in unità di $Mc^2$. Infatti in queste unità l'energia è uguale a $\gamma$, e nella formula di Bethe si possono fare le sostituzioni: $\gamma^2 \beta^2 = \gamma^2 - 1$ e $\beta^2 = 1 - \frac{1}{\gamma^2}$ [v. voce *Particelle elementari, passaggio attraverso la materia*, di C. Castagnoli, in «Enciclopedia della Scienza e della Tecnica Mondadori-McGraw-Hill», IX, p. 544 (1980)].

[39] L'effetto di densità causa una saturazione dell'aumento relativistico del potere frenante ad alte energie. Introducendo la correzione che tiene conto di tale effetto, descritta da $\frac{\delta}{2} = \ln \frac{\hbar \omega_p \beta \gamma}{I} - \frac{1}{2}$, $\hbar \omega_p \approx 137 \sqrt{4\pi \mathcal{N} r^3} \cdot mc^2$, la formula di Bethe assume la forma $\mathcal{F}_{Z_1} = \frac{4\pi mc^2 r_e^2 Z_1^2 \mathcal{N}}{\beta^2} \left\{ \ln\left(\frac{2mc^2 \gamma^2 \beta^2}{\overline{I}}\right) - \beta^2 - \frac{\delta}{2} \right\}$.

[40] A velocità ancor più basse la formula è inapplicabile (v. nota[34]).

[41] Si misura spesso la perdita d'energia in $MeV/g \cdot cm^{-2}$, anziché in $MeV/cm$, come il rapporto tra il potere frenante e la densità $\rho$ del mezzo (per l'aria $\rho \simeq 1,293 \, mg/cm^3$).

[42] Al campo magnetico terrestre contribuiscono, oltre al dipolo geocentrico che fornisce il contributo principale, termini minori di importanza decrescente (quadrupolo, ottupolo, ecc.), che provengono dallo sviluppo in serie del potenziale scalare magnetico $\Psi$. Limitatamente al solo termine dipolare si può considerare il campo magnetico terrestre pressappoco simmetrico rispetto all'asse di rotazione del nostro pianeta, con una componente radiale $B_r = 2\mu \cos\theta / r^3$ e una latitudinale $B_\theta = \mu \sin\theta / r^3$. Questo modello fornisce per il campo superficiale all'equatore ($\theta = 90^0$) il valore $B_r \simeq 0, B_\theta \simeq 31 \mu Tesla \simeq 0,31 gauss$, contro un valore misurato di $\simeq 24 \mu Tesla$ ($1 Tesla = 10^4 gauss$), e ai poli ($\theta = 0^0$) $B_\theta \simeq 0, B_r \simeq 62 \mu Tesla \simeq 0,62 gauss$, contro $\simeq 66 \mu Tesla$ misurati. In analogia con l'energia potenziale $e \cdot V$ di una carica in un campo elettrostatico, l'energia potenziale di un monopolo nel campo magnetico terrestre è $g_0 \cdot \Psi \sim g_0 \mu \cos\theta / r^2 \sim 68,5 e\mu \cos\theta / r^2 \sim 5 \cdot 10^9 \, eV$.

[43] La traiettoria di un monopolo in un campo magnetico uniforme, quale si può considerare in prima approssimazione quello terrestre, è uguale a quella di una particella carica in un campo elettrico. Scegliendo l'asse $z$ nella direzione del campo magnetico, la traiettoria giace su un piano che chiameremo piano $zs$, dove $s$ è un asse che giace nel piano $xy$ perpendicolare a $z$. Se scegliamo come istante iniziale $t = 0$ quello nel quale il monopolo ha una velocità nulla $v_z = 0$ lungo l'asse $z$ e passa per l'origine $z = s = 0$, si ricava facilmente che a bassa velocità ($v \ll c$) la traiettoria del monopolo è una parabola $z \simeq \frac{gH}{2mv_0^2} s^2$, dove $v_0$ è la velocità iniziale (interamente diretta lungo $s$) e $s$ è la lunghezza della proiezione della traiettoria nel piano $xy$ ($s = \sqrt{x^2 + y^2}$). Con più generali condizioni iniziali l'equazione della traiettoria diventa $z \simeq \frac{gH}{2mv_{s_0}^2} s^2 + \frac{v_{z_0}}{v_{s_0}} s + z_0$, con $s = \sqrt{(x-x_0)^2 + (y-y_0)^2}$. Nel caso relativistico si dimostra che la traiettoria è una catenaria $z = \frac{\gamma_0 mc^2}{gH} \text{ch} \frac{gH}{\gamma_0 mv_0 c} s$ [v. L.D. Landau - E.M. Lifchitz, *Théorie des champs*, MIR, Mosca (1970)].

[44] Si ricava da $g_0 \approx 68,5 e \, [u.e.s.]$ con $e \simeq 4,803 \cdot 10^{-10} \, [u.e.s.]$ applicando i fattori di conversione $1 u.e.s. = 1 erg / gauss \cdot cm$, $1 erg \simeq 6,242 \cdot 10^{11} eV$.

[45] Per una rassegna di questi sviluppi vedi: Sumil Mukhi, *Dirac's Conception of the Magnetic Monopole and its Modern Avatars*, in «Resonance», Vol.8, N.8 (August, 2003).